\documentclass[aps,prb,twocolumn,showpacs,superscriptaddress]{revtex4-1}
\usepackage{graphics}
\usepackage{subfigure}
\usepackage{longtable}
\usepackage{graphicx}
\usepackage{pstricks}
\usepackage{amsmath}
\usepackage{dcolumn}
\usepackage{bbm}
\usepackage{pdfpages}
\usepackage{amssymb}
\usepackage{subfigure}
\usepackage{epstopdf}
\usepackage{color}

\begin{document}

\title{\textit{Ab initio} determination of spin Hamiltonians with anisotropic exchange interactions: 
the case of the pyrochlore ferromagnet Lu$_2$V$_2$O$_7$}

\author{Kira Riedl}
\email{riedl@itp.uni-frankfurt.de}
\affiliation{Institut f\"ur Theoretische Physik, Goethe-Universit\"at Frankfurt, 
Max-von-Laue-Stra{\ss}e 1, 60438 Frankfurt am Main, Germany}

\author{Daniel Guterding}
\affiliation{Institut f\"ur Theoretische Physik, Goethe-Universit\"at Frankfurt, 
Max-von-Laue-Stra{\ss}e 1, 60438 Frankfurt am Main, Germany}

\author{Harald O. Jeschke}
\affiliation{Institut f\"ur Theoretische Physik, Goethe-Universit\"at Frankfurt, 
Max-von-Laue-Stra{\ss}e 1, 60438 Frankfurt am Main, Germany}

\author{Michel J. P. Gingras}
\affiliation{Department of Physics and Astronomy, University of Waterloo, Ontario, N2L 3G1, Canada}
\affiliation{Perimeter Institute for Theoretical Physics, Waterloo, Ontario, N2L 2Y5, Canada}
\affiliation{Canadian Institute for Advanced Research, 180 Dundas Street West, Suite 1400, Toronto, ON, M5G 1Z8, Canada}

\author{Roser Valent\'i}
\affiliation{Institut f\"ur Theoretische Physik, Goethe-Universit\"at Frankfurt, 
Max-von-Laue-Stra{\ss}e 1, 60438 Frankfurt am Main, Germany}

\begin{abstract}

  We present a general framework for deriving effective spin
  Hamiltonians of correlated magnetic systems based on a combination
  of relativistic {\it ab initio} density functional theory
  calculations (DFT), exact diagonalization of a generalized Hubbard
  Hamiltonian on finite clusters and spin projections onto the
  low-energy subspace. A key motivation is to determine anisotropic bilinear
  exchange couplings in materials of interest. As an example, we apply this method to the
  pyrochlore Lu$_2$V$_2$O$_7$ where the vanadium ions form a lattice
  of corner-sharing spin-1/2 tetrahedra. In this compound, anisotropic Dzyaloshinskii-Moriya interactions
  (DMI) play an essential role in inducing a magnon Hall effect. We obtain quantitative
  estimates of the nearest-neighbor Heisenberg exchange, the
  DMI, and the symmetric part of
  the anisotropic exchange tensor. Finally, we compare our results
  with experimental ones on the Lu$_2$V$_2$O$_7$ compound.
 
\end{abstract}


\maketitle

\section{Introduction}
The Heisenberg Hamiltonian and its extensions are among the most
successful models for describing magnetism in correlated
systems~\cite{Fazekas1999,Khomskii2010}. However, for an accurate
description of real material properties, a sound understanding of the
role of the lattice structure (e.g. superexchange pathways) and its
consequence on the spin-spin exchange parameters is indispensable.
Various methods for determining exchange parameters for real materials
exist. A popular one consists of fitting calculated properties
obtained by assuming a particular form
of the spin Hamiltonian to experimental data (specific heat, magnetic
susceptibility, magnetization, inelastic neutron scattering, {\it
  etc.})~\cite{Coldea2002,Chaboussant2004,Silverstein2014,Savary2012}.
A complementary procedure that is gaining popularity is to estimate
the coupling constants of the Heisenberg Hamiltonian from methods
based on first principles, like mapping total energies obtained from
spin-polarized density functional theory (DFT) calculations to a
Heisenberg model~\cite{Xiang2013,Janson2008,Jeschke2011,Foyevtsova2011,Jeschke2013,Tutsch2014}.  Both
approaches are useful for providing information on Heisenberg-only
interactions.  These become, however, problematic when terms other
than rotationally-invariant (isotropic) Heisenberg exchange
$\mathcal{J}_{ij} (\vec{S}_i \cdot \vec{S}_j)$ are not negligible, as
it happens in rare-earth pyrochlore compounds~\cite{YbTO_contro}
or, even when small, they play a crucial role in the physics of the
system~\cite{Sadeghi2015,ring_exchange}.  Common examples of such
anisotropic couplings are the off-diagonal Dzyaloshinskii-Moriya
vector $\vec{\mathcal{D}}_{ij}$ and the traceless symmetric tensor
$\hat{\mathcal{K}}_{ij}$:
\begin{align} \label{eq:HSpin}
H_{\text{spin}}
&=\mathcal{J}_{ij} \, (\vec{S}_i \cdot \vec{S}_j) + \vec{\mathcal{D}}_{ij} \cdot ( \vec{S}_i \times \vec{S}_j ) +
\vec{S}_i \cdot \hat{\mathcal{K}}_{ij} \cdot \vec{S}_j . 
\end{align}
%
By broadly aiming to obtain reliable quantitative estimates of the coupling
constants in a general spin Hamiltonian such as that of
Eq.~\eqref{eq:HSpin}, we explore here, as a first motivation for our
work, a method that combines DFT calculations with exact
diagonalization of the electronic (Hubbard-like) Hamiltonian on finite
clusters.  This approach does not depend on experimental input, except
for the crystal structure.
\begin{figure}
 \includegraphics[width=0.5\textwidth]{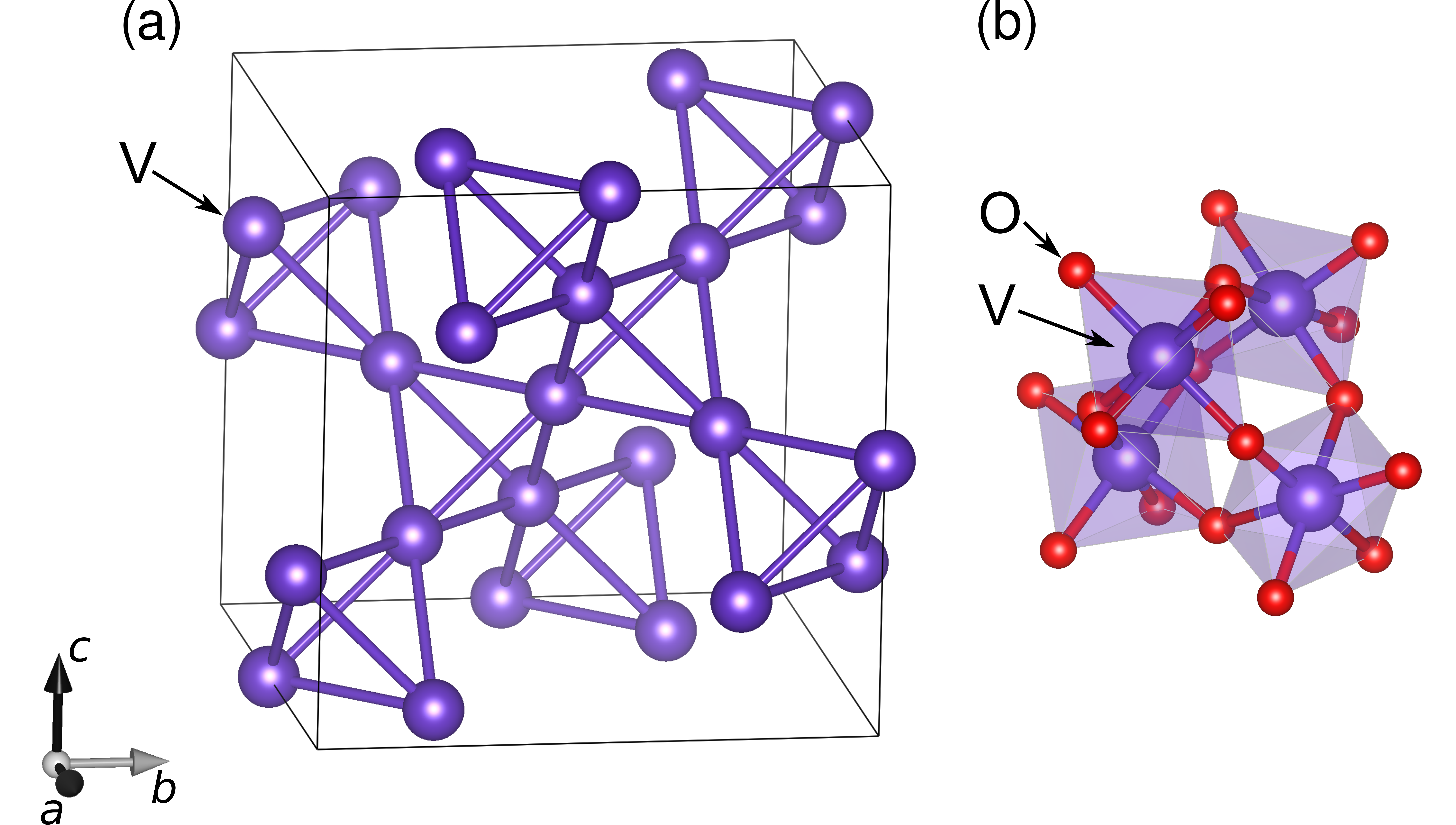}
  \caption{(a) Network of corner-sharing vanadium tetrahedra in 
    the pyrochlore Lu$_2$V$_2$O$_7$. (b) Oxygen environment around a vanadium tetrahedron.}
 \label{fig:crystal_struc}
\end{figure}

As a specific application of the method, we evaluate the bilinear
spin-spin coupling constants in Eq.~\eqref{eq:HSpin} of the insulating
Lu$_2$V$_2$O$_7$ pyrochlore ferromagnet. This material has recently
been proposed as a candidate topological magnon insulator with chiral
edge states~\cite{Zhang2013,Mook2014} and evidence for a magnon Hall effect has
also been reported~\cite{Onose2010}.  The magnetic properties of
Lu$_2$V$_2$O$_7$ are dominated by corner-sharing spin-1/2 vanadium
tetrahedra (see Fig.~\ref{fig:crystal_struc}).  Due to the lack of
bond-inversion symmetry for the pyrochlore lattice\cite{Elhajal2005}, the
Dzyaloshinskii-Moriya interaction (DMI) of spin-orbit
origin~\cite{Moriya1960, Dzyaloshinskii1958} may not be negligible.
The DMI is expected to play an essential role on the observed magnon
Hall effect in Lu$_2$V$_2$O$_7$ and there is a debate about the
magnitude of the principal spin-spin interactions in this material.
The need to better understand the scale of the anisotropic
interactions in contemporary magnetic systems and their role on
topological magnon transport is further emphasized by the observation
of such phenomena in materials~\cite{Hirschberger,Chisnell} other than
Lu$_2$V$_2$O$_7$.  From a broader context, the latter material may
then possibly be viewed as an important test bench for establishing
close contact between theory and experiment.
 
Coming back to Lu$_2$V$_2$O$_7$, experimental and theoretical results
have been reported for the nearest-neighbor Heisenberg exchange,
$\mathcal{J}_{ij}$, and the Dzyaloshinskii-Moriya vector,
$\vec{\mathcal{D}}_{ij}$.  However, no consensus has yet emerged on
the value of these two spin-spin couplings.  Fitting transport and
magnetic specific heat data on Lu$_2$V$_2$O$_7$~\cite{Onose2010} leads
to ferromagnetic (negative) $\mathcal{J}_{ij} \simeq
-3.4$~meV~\cite{stiffness} and $\vert\vec{\mathcal{D}}_{ij} \vert /
\vert \mathcal{J}_{ij} \vert \simeq 1/3$.  In contrast, recent
inelastic neutron scattering measurements\cite{Mena2014} indicate that
$\vert \vec{\mathcal{D}}_{ij} \vert / \vert \mathcal{J}_{ij} \vert
\simeq 0.18$.  On the other hand, Xiang {\it et al.}~\cite{Xiang2011}
obtained $\vert \vec{\mathcal{D}}_{ij} \vert / \vert \mathcal{J}_{ij}
\vert=0.048$ by mapping DFT total energies to a spin Hamiltonian. This
ratio is one order of magnitude smaller than the values obtained from
fitting to experimental data. However, one should note that
Ref.~\onlinecite{Xiang2011} includes an additional single-ion
anisotropy term in the effective spin Hamiltonian used to parameterize
the energy of magnetic moment configurations.  In a quantum spin-1/2
Hamiltonian, any such even-power term should, however, be absent as
they are trivially proportional to the identity (Pauli, $\sigma_0$)
matrix.
Considering the disparities between the values so far determined for
$\mathcal{J}_{ij}$ and $\vec{\mathcal{D}}_{ij}$, one is naturally led
to ask whether additional symmetry-allowed terms, like the
symmetric tensor $\hat{\mathcal{K}}_{ij}$ in Eq.~\eqref{eq:HSpin}, 
are truly negligible in this compound. 
It is therefore of some importance to determine such couplings
consistently within a well-defined calculational procedure; this is
the second main motivation for our work.

The paper is organized as follows: Section~\ref{HAMIL} discusses the various
steps necessary to establish the generalized spin Hamiltonian that we
seek.  We first present the framework for obtaining tight-binding
parameters and the spin-orbit coupling constant $\lambda$ out of
relativistic DFT calculations. In a second step, we perform an exact
diagonalization of a generalized Hubbard Hamiltonian that includes the
\textit{ab initio} tight-binding parameters and $\lambda$.
Introducing effective spin-1/2 operators, we project the results on
the low-energy subspace of the system to extract an effective spin
Hamiltonian, which allows us to determine the various exchange coupling
constants. The method is applied to Lu$_2$V$_2$O$_7$ where we compare
briefly to experimental results.  We conclude the paper with a summary
in Section~\ref{DISCUSSION}.

\section{Generalized Model Hamiltonian}\label{HAMIL}

\subsection{{\it Ab initio} determination of the tight-binding hopping and spin-orbit parameters} 

Our starting Hamiltonian is a generalized multiorbital Hubbard model for $d$ electrons that includes spin-orbit coupling (SOC):
\begin{align} \label{eq:H_general}
H_{\text{tot}} = H_{\text{hop}} + H_{\text{soc}} + H_{\text{int}},
\end{align}
where
\begin{align} \label{eq:H_hop}
H_{\text{hop}}=& \sum_{ ij } \sum_{\alpha \beta} \, t_{i \alpha, j \beta} \, d_{i \alpha }^\dagger d_{j \beta }
\end{align}
is the hopping term with hopping parameters $t_{i \alpha, j \beta}$
where $i,j$ are site indices and $\alpha,\beta$ are orbital indices
\cite{Footnote:orbidx}.
\begin{align} \label{eq:H_SOC}
H_{\text{soc}}=\lambda \sum_{i} \sum_{\alpha \beta} \sum_{\sigma \sigma^\prime} \langle \, i \, \alpha \, \sigma \vert \vec{L} \cdot \vec{S} \vert \, i \, \beta \, \sigma^\prime \rangle \, d_{i \alpha \sigma}^\dagger d_{i \beta \sigma^\prime} ,
\end{align}
is the spin-orbit term where $\lambda$ is the strength of the on-site
spin-orbit coupling and $\sigma$ and $\sigma'$ are the spin indices.
\begin{align}\label{H_int} 
H_{\text{int}}=& \, \sum_i \sum_{\alpha \beta} U_{\alpha \beta} n_{i \alpha \uparrow} n_{i \beta \downarrow} \nonumber \\
&+ \frac{1}{2} \sum_{i \sigma} \sum_{\alpha \neq \beta} ( U_{\alpha \beta} - J_{\alpha \beta} ) n_{i \alpha \sigma} n_{i \beta \sigma} \nonumber \\
&+ \sum_i \sum_{\alpha \neq \beta} J_{\alpha \beta} ( d_{i \alpha \uparrow}^\dagger d_{i \beta \downarrow}^\dagger d_{i \alpha \downarrow} d_{i \beta \uparrow} + d_{i\alpha \uparrow}^\dagger d_{i \alpha \downarrow}^\dagger d_{i\beta \downarrow} d_{i \beta \uparrow})
\end{align}
is the two-particle interaction term for $3d$
electrons~\cite{Pavarini2011}.  There are two independent parameters
in this Hamiltonian, the Coulomb repulsion of electrons on the same
orbital, $U_0$, and the average Hund's coupling,
$J_{\text{avg}}=\frac{1}{2l(2l+1)}\sum_{\alpha \neq \beta} J_{\alpha
  \beta}$, with $U_{\alpha \beta}=3 \, U_0 \mathbbm{1} - 2 J_{\alpha
  \beta}$.  The explicit form of the interaction matrices $U_{\alpha
  \beta}$ and $J_{\alpha \beta}$ is given in Appendix
\ref{sec:H_int_parameters}.

We first determine via {\it ab initio} methods the hopping parameters
$t_{i \alpha, j \beta}$ in Eq.~\eqref{eq:H_hop} and then the
spin-orbit coupling constant $\lambda$ in Eq.~\eqref{eq:H_SOC}. We
perform non-magnetic, non-relativistic DFT calculations within an
all-electron full-potential local orbital (FPLO)\cite{FPLOmethod}
basis and use for the exchange-correlation functional the generalized
gradient approximation (GGA)\cite{PerdewBurkeErnzerhof}.  The hopping
parameters are then obtained from projective Wannier
functions~\cite{Altmeyer2015,Guterding2015} as implemented in
FPLO\cite{FPLOtightbinding}.

\begin{figure} 
\includegraphics[width=0.49\textwidth]{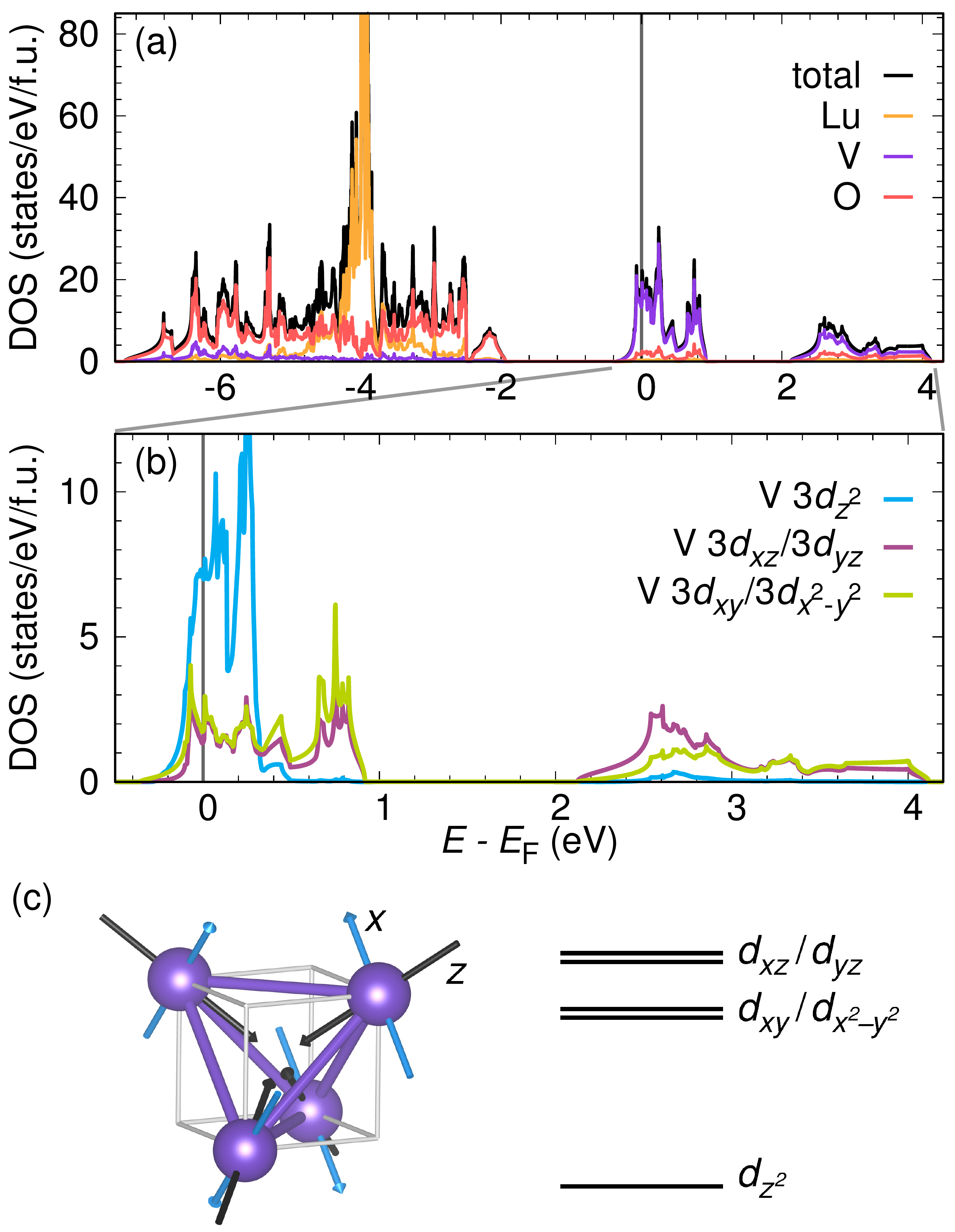} 
\caption{(a) Non-relativistic density of states (DOS) of
  Lu$_2$V$_2$O$_7$ in the energy range $[-7\,\text{eV},4\,\text{eV}]$
  obtained in the GGA approximation.  Shown are the total, as well as
  partial DOS corresponding to V, O and Lu. (b) Vanadium
  orbital-resolved DOS around the Fermi level. (c) Illustration of the
  local reference frame in one tetrahedron of vanadium atoms, and
  orbital energy hierarchy in Lu$_2$V$_2$O$_7$.}
  \label{fig:DOS+lcoordinates+orbitals}
\end{figure}

For the Lu$_2$V$_2$O$_7$ pyrochlore, we use the experimental structure
determined by Haghighirad {\it et al.}~\cite{Haghighirad2011}.  We
show in Fig.~\ref{fig:DOS+lcoordinates+orbitals} the total density of
states which is dominated by vanadium $3d$ weights near the Fermi
level.  Because of the distorted oxygen octahedra surrounding each
vanadium atom, illustrated in Fig.~\ref{fig:crystal_struc}~(b), there
is a trigonal crystal field splitting of the $d$ orbitals (see right
panel of Fig.~\ref{fig:DOS+lcoordinates+orbitals}~(c)).  This results
in doubly-degenerate $d_{xy}$ and $d_{x^2\text{-}y^2}$, as well as
$d_{xz}$ and $d_{yz}$ orbitals. Our choice of local coordinate systems
at each vanadium ion is the same as the one used in
Ref.~\onlinecite{Ross2011}. The $z$ axes point along the cubic
$\langle 111 \rangle$ directions, the $x$ axes point along the cubic
$\langle 0 1 \bar{1} \rangle$ directions while the $y$ axes point
along the $\langle \bar{2} 1 1\rangle$ directions such that $\langle
xyz \rangle$ form a local orthogonal triad, as illustrated in
Fig.~\ref{fig:DOS+lcoordinates+orbitals}~(c).  For a detailed
description, we list in Appendix~\ref{sec:local_coordinates} the most
relevant onsite (see Table~\ref{tab:Hop_onsite}) and nearest-neighbor
(see Table~\ref{tab:Hop_NN}) hopping parameters.

\begin{figure} 
\includegraphics[width=0.45\textwidth]{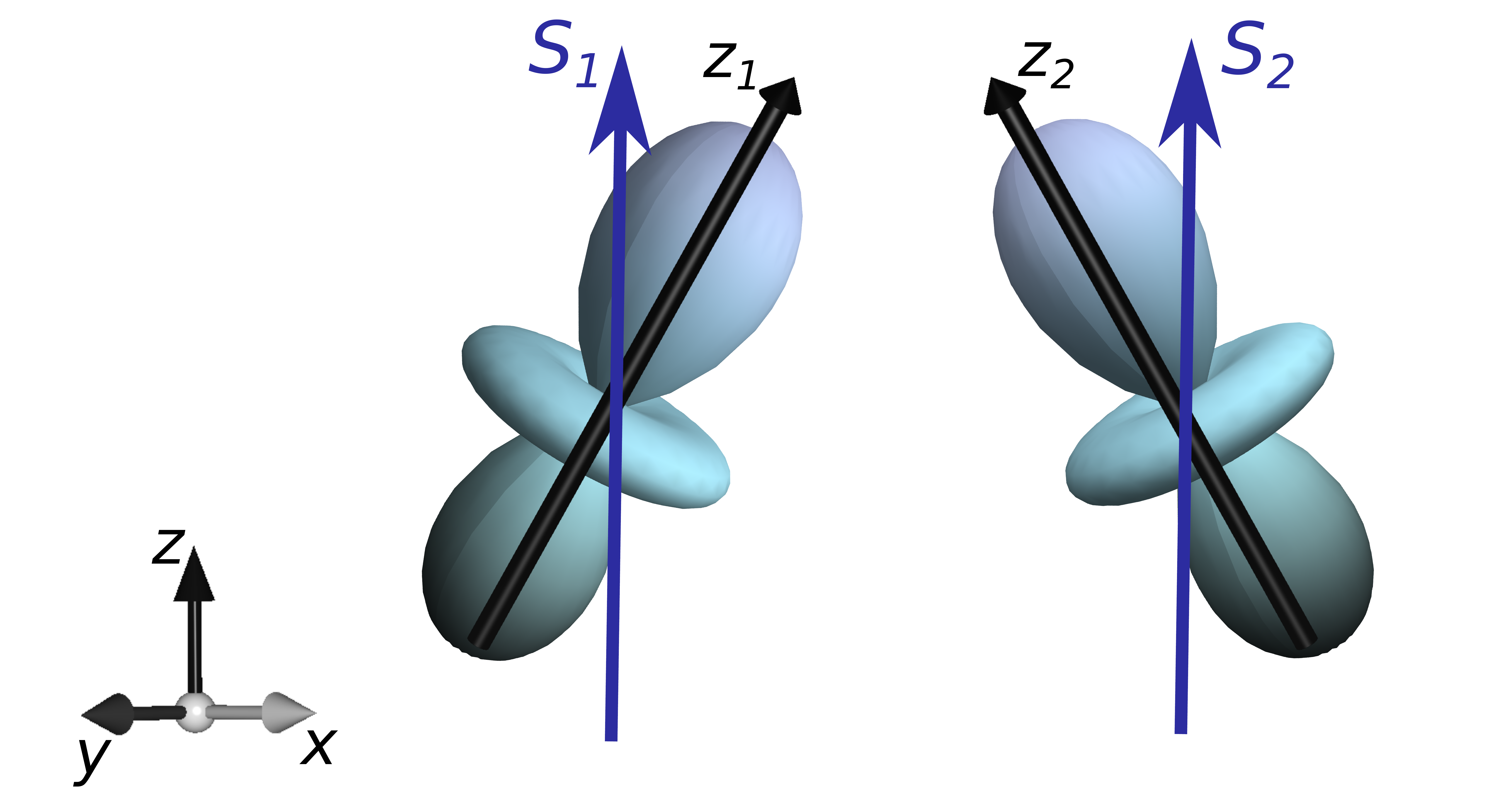} 
\caption{Reference frames of spin and orbital degrees of freedom
on two neighboring sites.
The spins S$_1$, S$_2$ are given in the global reference frame
and the orbitals are $d_{z^2}$ orbitals in the local reference frame
at each site.  Due to the site-dependent local coordinate
  frame, the spin operators have to be rotated in each local reference
frame.} \label{fig:LS_scalar_product}
\end{figure}

Having determined $t_{i \alpha, j \beta}$, we next proceed to compute
$\lambda$. We first derive the analytical expressions for the
spin-orbit coupling matrix elements. The scalar product $\vec{L} \cdot
\vec{S} = \sum_{r_i} L_{r_i} S_{r_i}$ leads to a dependence on the
direction of the local axes $r_i=\{x_i,y_i,z_i\}$ at site $i$. The
matrix elements can be evaluated using the Kronecker product,
\begin{align}\label{hamil_ls}
\sum_{{r}_i} \langle i \alpha \, \sigma \vert (L_{{r}_i} S_{{r}_i}) \vert  i \beta \, \sigma^\prime \rangle 
= \sum_{{r}_i} \langle \alpha_i  \vert L_{r_i} \vert  \beta_i \rangle \otimes \langle \sigma  \vert S_{r_i}
 \vert \sigma^\prime \rangle,
\end{align}
where $\alpha_i$, $\beta_i$ label the site dependent $d$ orbitals at
site $i$.  The spin operator $S_{r_i}$ should have the components aligned
along the local
coordinate frame while the state $\vert \sigma^\prime \rangle=\{\vert \uparrow \rangle, \vert \downarrow \rangle \}$ is
defined in a global coordinate frame. We therefore have to rotate the
spin operator in each local reference frame (see
Fig.~\ref{fig:LS_scalar_product}).  For example, for site No. 1, the
local $z$ axis, $\vec {z}_1$, expressed in the global coordinate
system is
\begin{align*}
\vec{z}_1=\frac{1}{\sqrt{3}}\begin{pmatrix}
  1 \\ 1 \\ 1
\end{pmatrix}.
\end{align*}
Therefore, the spin operator measuring the local $z$ component at this
site is
\begin{align*}
S_{z_1}=\frac{1}{\sqrt{3}}(S_x + S_y + S_z ),
\end{align*}
where the spin operators are $\vec{S}=\frac{1}{2} \vec{\sigma}$, with
$\vec{\sigma}$ the Pauli matrices.  On the other hand, the matrix
elements of the angular momentum are evaluated at each vanadium local
coordinate frame as given in Eq.~\eqref{hamil_ls}.

By application of the Kronecker product in Eq.~\eqref{hamil_ls}, the
analytical expressions for the spin-orbit coupling matrix elements at
every site are obtained, leaving only the spin-orbit coupling strength
$\lambda$ to be determined. Two main properties contribute to the value of
$\lambda$: the nature of the ion (vanadium V$^{4+}$ here) for which the
spin-orbit interaction is being considered and, to a smaller degree,
the crystal environment. In order to take into account these effects,
we perform fully relativistic DFT calculations with FPLO and map via a
numerical fitting procedure the sum of H$_{\text{hop}}$ in
Eq.~\eqref{eq:H_hop} and H$_{\text{SOC}}$ in Eq.~\eqref{eq:H_SOC},
where H$_{\text{hop}}$ contains the hopping parameters previously
determined to the relativistic DFT bandstructure. The only parameter
left to fit the relativistic DFT bandstructure is $\lambda$.

\begin{figure}
\includegraphics[width=0.45\textwidth]{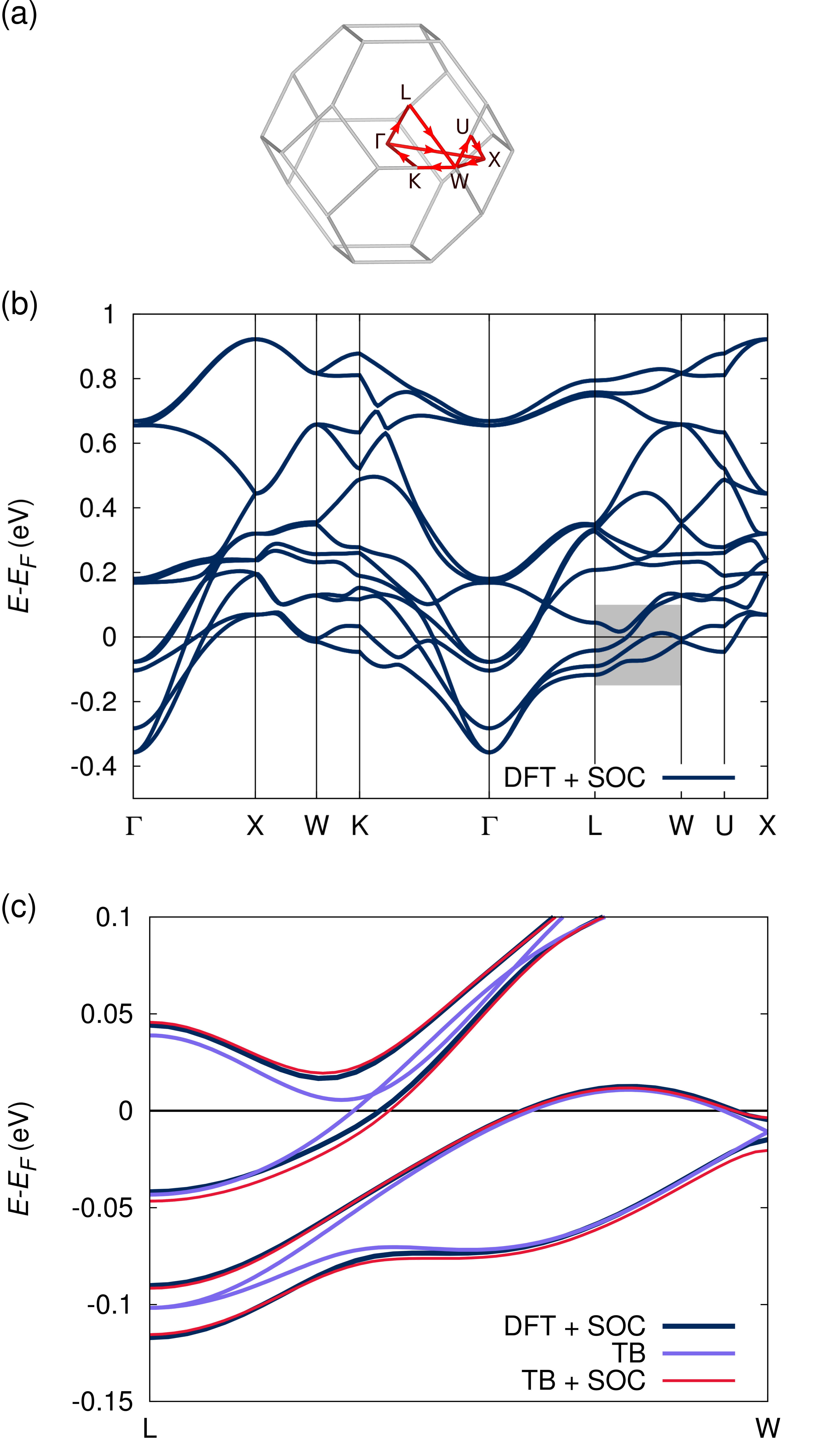}
\caption{(a) Chosen high-symmetry \boldmath{$k$}-path in the Brillouin
  zone of pyrochlore.  (b) Relativistic band structure of
  Lu$_2$V$_2$O$_7$ on the high-symmetry path. (c) Relativistic band
  structure between the high symmetry points L and W in a smaller
  energy window around the Fermi energy (gray shaded region in
  (b)). The dark blue (DFT$+$SOC) curve is the result of the fully
  relativistic band structure calculation. The purple curve (TB)
  represents the tight-binding band structure from the
  non-relativistic calculation. The red (TB$+$SOC) curve is the result
  of the tight-binding band structure taking the spin-orbit coupling
  term into account. The band splitting caused by relativistic effects
  is well reproduced, as can be seen by comparing the blue (DFT$+$SOC)
  and red (TB$+$SOC) curves.}
\label{fig:tb_bandstructure}
\end{figure}

We illustrate this procedure in Fig.~\ref{fig:tb_bandstructure} for
Lu$_2$V$_2$O$_7$. As expected, the spin-orbit coupling causes band
splittings with respect to the non-relativistic band structure:
compare the tight-binding band structure represented by the purple
curve (which reproduces well the non-relativistic DFT band
structure; not shown), with the fully relativistic band structure
calculation, given by the blue curve in
Fig.~\ref{fig:tb_bandstructure}~(c).  Including the spin-orbit
coupling contribution in the model Hamiltonian leads to a good
representation of the relativistic band structure (red curve in
Fig.~\ref{fig:tb_bandstructure}~(c)) from which we can extract
$\lambda$ through optimization. For Lu$_2$V$_2$O$_7$, we find $\lambda
\sim 30.0$~meV.  As a reference, we note that
$\lambda_{\text{exp}} = 30.75$~meV for an isolated vanadium
(V) atom~\cite{BlumeWatson1963}.

\subsection{Cluster diagonalization of H$_{\text{tot}}$}

At this point, having determined from {\it ab initio} calculations
$t_{i \alpha, j \beta}$ and $\lambda$, we are only left with the
interaction parameters $U_0$ and $J_{\text{avg}}$ in H$_{\text{int}}$, given in
Eq.~\eqref{H_int}. These two values will be left as model parameters
and we will discuss them further below.  Our aim here is to obtain a
low-energy spin Hamiltonian out of the generalized Hubbard Hamiltonian
H$_{\text{tot}}$ Eq.~\eqref{eq:H_general}. To this effect, we proceed with a
cluster diagonalization of H$_{\text{tot}}$, focusing on the
2-site case in the example of Lu$_2$V$_2$O$_7$.

We note the importance of the Hund's coupling for this multiorbital
system.  In Lu$_2$V$_2$O$_7$, the ground state is ferromagnetic
\cite{Shamoto2002}. The angle between two vanadium atoms and the
nearest oxygen atom is $\theta=131.44^{\circ}$. This is neither close
to $180^{\circ}$, where according to the Goodenough-Kanamori rules
antiferromagnetic coupling is favored, nor to $90^{\circ}$ which would
lead to ferromagnetic coupling~\cite{Khomskii2014}.  Miyahara {\it et
  al.} \cite{Miyahara2007} suggested that ferromagnetism in
Lu$_2$V$_2$O$_7$ is induced by orbital ordering.  These authors argue
that the orbital ordering in Lu$_2$V$_2$O$_7$ leads to such a large
ratio of hopping amplitudes $t_{i z^2, j xy}/t_{i z^2,j z^2}$ and
$t_{i z^2, j x^2\text{-}y^2}/t_{i z^2, j z^2}$ that a ferromagnetic
ground state is induced. This is of course only possible when a
mechanism exists that favors ferromagnetic arrangements on different
orbitals, the Hund's coupling. We note that the interaction part used
in Ref.~\onlinecite{Miyahara2007} is a simplified version of the
correct $3d$ Hubbard Hamiltonian\cite{Pavarini2011}.  Nevertheless,
these arguments suggest that it is not possible to neglect the various
Hund's couplings, $J_{\alpha \beta}$, in the Hamiltonian and, at the
same time, reproduce the correct ferromagnetic V$-$V exchange.

Notwithstanding the importance of considering the $J_{\alpha \beta}$
couplings, there is a critical reason why
all five V 3$d$ orbitals need to be included in the calculations.  The
oxygen octahedra surrounding the vanadium atoms are slightly
distorted. This induces a trigonal crystal field splitting of the $d$
orbitals (see Fig.~\ref{fig:DOS+lcoordinates+orbitals}), with the
lowest level being non-degenerate.  The necessity of including all
five $d$ orbitals is made evident by invoking simple
perturbation-theory considerations.  Specifically, the importance of
the various states to the effective spin Hamiltonian that we aim to
determine scales roughly with the inverse of the crystal field
splitting. In a hypothetical case where the lowest orbitals were
degenerate, these would play the main role in the physics of the
system and it would be justified to consider only those. Otherwise,
one has to take all orbitals into account. Furthermore, it is
important to note that the difference of the magnetic quantum numbers
of the lowest orbital $d_{z^2}$ ($m_l=0$) with the next higher
orbitals $d_{xy}/d_{x^2\text{-}y^2}$ ($m_l=\pm2$) is two.  As a
result, the spin-orbit coupling $\vec{L} \cdot \vec{S} =
\frac{1}{2}(L_- S_+ + L_+ S_-) + L_z S_z$ has almost no contribution
if we neglect the two highest energy orbitals, $d_{xz}/d_{yz}$ with
$m_l=\pm1$ (see Fig.~\ref{fig:DOS+lcoordinates+orbitals}~(c)).

H$_{\text{tot}}$ is diagonalized for two sites, five $d$ orbitals and
two spin degrees of freedom. The filling counting in Lu$_2$V$_2$O$_7$
is one electron per $V$ site so that we constrain the subspace to
states containing two vanadium ions. In second quantization, the
two-site/two-particle system has ${{20}\choose{2}}=190$ states, and we
therefore need to diagonalize a $190 \times 190$ matrix.  Note that we
have within this approach the constraint $U_0-3 J_{\alpha \beta}>0$
for all orbitals; otherwise, states with two particles on a single
site would become favorable because the system then gains energy with
double occupation, and the projection onto singly-occupied states is
no longer justified.
 
In the following paragraphs, we investigate the properties of the four
singly-occupied states and their corresponding energy. We define the
low energy state $\vert \psi \rangle$ via a linear combination of
singly occupied states with coefficients $c_{\sigma\sigma'}$
($\sigma=\uparrow,\downarrow$)
\begin{align} \label{eq:state}
\vert \psi \rangle = & \; 
  c_{\uparrow \uparrow}     \, \vert  \! \uparrow_{i,z^2} \uparrow_{j,z^2} \rangle 
+ c_{\uparrow \downarrow}   \, \vert  \! \uparrow_{i,z^2} \downarrow_{j,z^2} \rangle \nonumber \\
&+ c_{\downarrow \uparrow}  \, \vert  \! \downarrow_{i,z^2} \uparrow_{j,z^2} \rangle 
+ c_{\downarrow \downarrow} \, \vert  \! \downarrow_{i,z^2} \downarrow_{j,z^2} \rangle
\end{align} 
where $i,j$ are site indices.

We first discuss various limiting cases.  In the nonrelativistic
atomic limit ($\lambda$=0 and all hopping terms set to zero), the
ground state is four-fold degenerate with every site being singly
occupied with the electron located in the orbital of lowest energy; in
Lu$_2$V$_2$O$_7$ this is the $d_{z^2}$ orbital.  For two sites, the
ground state energy is then $\varepsilon_0=2\varepsilon_{z^2}$ which
is twice the on-site energy of the $d_{z^2}$ orbital.

If we switch on spin-orbit coupling, states with certain orbitals and
spins get admixed, and the eigenenergies undergo a shift in the atomic
limit.  Henceforth, we deal with pseudo-orbitals $\tilde{\alpha}$ with
an energy for the lowest state
$\varepsilon^{\text{SO}}_{\tilde{z}^2}$.
 The ground state energy of the two-site system is then twice
the on-site energy of the pseudo-orbital with the lowest energy
$\varepsilon_0^{\text{SO}}=2 \varepsilon_{\tilde{z}^2}^{\text{SO}}$.

\begin{table*}
\begin{tabular}{p{2cm}|p{2cm}|p{2cm}|p{2cm}|p{2cm}}
$\varepsilon^{\text{hop}}$ (eV) & $c_{\uparrow \uparrow}$	& $c_{\uparrow \downarrow}$ & $c_{\downarrow \uparrow}$	& $c_{\downarrow \downarrow}$ \\
\hline
\hline
0.43907 & 1 & 0	& 0	& 0 \\ 
\hline
0.43907 & 0 & 0 & 0 & 1 \\ 
\hline
0.43907 & 0 & 0.70 & 0.70 & 0 \\ 
\hline
0.44769 & 0 & 0.70 & $-0.70$ & 0
\end{tabular}

\bigskip
\begin{tabular}{p{2cm}|p{2cm}|p{2cm}|p{2cm}|p{2cm}}
$\varepsilon^{\text{SO+hop}}$ (eV) & $c_{\uparrow \uparrow}$	& $c_{\uparrow \downarrow}$ & $c_{\downarrow \uparrow}$	& $c_{\downarrow \downarrow}$  \\
\hline
\hline
0.43300 & $0.69-0.06i$ & 0	& 0	& $0.69-0.06i$ \\ 
\hline
0.43306 & $-0.47+0.15i$ & $0.47-0.15i$ & $0.47-0.15i$ & $0.47-0.15i$ \\ 
\hline
0.43307 & $0.49-0.02i$ & $0.49$ & $0.49-0.05i$ & $-0.49-0.02i$ \\ 
\hline
0.44104 & $0.02i$ & $0.69-0.03i$ & $-0.69+0.06i$	& $0.02i$	 
\end{tabular}
\caption{Coefficients of the low energy states in Lu$_2$V$_2$O$_7$: 
  (a) Without spin-orbit coupling, there are two energy levels, one of them triply degenerate with $\varepsilon^{\text{hop}}_0=0.43907$ eV, 
  and a singlet with 
  $\varepsilon^{\text{hop}}_1=0.44769$ eV. (b) Full 
  Hamiltonian, there are four distinct energy levels.}
\label{tab:Coeff2+3}
\end{table*}

If we switch on hopping, but neglect spin-orbit coupling, we observe a
triplet-singlet splitting in the energy spectrum and additional
contributions from states that are not the low energy states in the
atomic limit are admixed. Without the Hund's couplings, antisymmetric
states are energetically favored since the Pauli principle allows
enhanced hopping processes in this case. The Hund's coupling
$J_{\alpha \beta}$ represents a competing mechanism and can, depending
on its strength, lead to the symmetric states being lowest in
energy. Results for the energies $\varepsilon^{\text{hop}}$ in
Lu$_2$V$_2$O$_7$ are given in Table~\ref{tab:Coeff2+3}~(a) where we
chose $U_0=3.3$ eV and $J_{\text{avg}}=0.845$ eV.

By diagonalizing the full Hamiltonian, the mixing of orbitals and
spins due to spin-orbit coupling combined with the orbital-dependent
hoppings lead to an additional, very small splitting of the three
lowest states for Lu$_2$V$_2$O$_7$ given by the energies
$\varepsilon^{\text{SO+hop}}$ in Table~\ref{tab:Coeff2+3}~(b).  In
Lu$_2$V$_2$O$_7$ the lowest energies are between
$\varepsilon^{\text{SO+hop}}_0=0.433$~eV and
$\varepsilon^{\text{SO+hop}}_4=0.44104$~eV while the next higher
eigenenergy (not shown) is $\varepsilon^{\text{SO+hop}}_5=0.59827\,$eV.
This energy gap leads to a well-defined separation
 of the low energy states from the excited states which allows us
to focus on the low energy states in the analysis below.

\subsection{Effective Spin Hamiltonian}

With the detailed information on the low energy states of the 2-site
system now in hand, we construct an effective spin Hamiltonian acting
within the low-energy subspace given by the four states described
above. Within these four states, we neglect the very small
coefficients of basis states which do not describe singly-occupied
states in the low energy orbital.  In this way, we construct a basis
that is not orthonormal, $\vert b_j \rangle = \sum_i c_i \vert s_i
\rangle$, where $\vert s_i \rangle$ are the four singly-occupied low
energy states as in Eq.~(\ref{eq:state}).  The coefficients $c_i$ are
those of Table \ref{tab:Coeff2+3}~(b).  The overlap matrix $P$, with
elements $P_{ij}\equiv c_i^\ast c_j \langle s_i \vert s_j \rangle$ for
Lu$_2$V$_2$O$_7$, is diagonal with overlaps around 0.96.

After orthonormalization~\cite{Szczepanik2013}, the coefficients $c_i$
are slightly modified (shown in Table \ref{tab:Coeff_ortho}) while the
eigenvalues are unchanged.  At this point of the calculation, the
effective Hamiltonian is given in the orthonormalized basis $\vert b_j
\rangle$, with the coefficients given in Table~\ref{tab:Coeff_ortho}.

\begin{table*}
\begin{tabular}{p{1.5cm}|r|r|r|r}
$\varepsilon$ (eV) & $c^\prime_{\uparrow \uparrow}$	& $c^\prime_{\uparrow \downarrow}$ & $c^\prime_{\downarrow \uparrow}$	& $c^\prime_{\downarrow \downarrow}$   \\
\hline
\hline
0.43300  & $0.7043-0.0626i$ & 0 & 0 & $0.7043-0.0626i$ \\
\hline
0.43306 & $-0.4758+0.1526i$ & $0.4764-0.1529i$  & $0.4764-0.1528i$ & $0.4758-0.1526i$\\
\hline
0.43307  & $-0.4995+0.0245i$ & $0.4999$ &  $0.4975+0.0493i$ & $-0.4995-0.0245i$\\
\hline
0.44104  & $-0.0012-0.0175i$ &$-0.7063+0.0295i$ & $0.7039-0.0645i$ & $0.0012+0.0175i$
\end{tabular}
\caption{Coefficients of the states in the orthonormal low-energy for Lu$_2$V$_2$O$_7$.} \label{tab:Coeff_ortho}
\end{table*}

As an alternative approach for constructing the effective spin
Hamiltonian, as well as a check for consistency, we also performed
second order perturbation theory and compared the resulting effective
Hamiltonian with the one obtained via the cluster diagonalization. In
second order perturbation theory\cite{Lindgren1986},
$H_{\text{eff}}^{\text{PT}}=\mathbbm{P} \, H_{\text{hop}}^{i \neq j }
\, \mathbbm{R} \, H_{\text{hop}}^{i \neq j} \, \mathbbm{P}$ up to two
intersite hopping processes are considered. The operator
$\mathbbm{P}=\sum_i \vert s_i \rangle \langle s_i \vert$ projects onto
the low energy subspace while $\mathbbm{R}=\sum_{ij} \vert e_i \rangle
\langle e_i \vert ( \varepsilon_0 - H_0 )^{-1} \vert e_j \rangle
\langle e_j \vert$ projects onto the renormalized subspace of 
excited states $\vert e_i \rangle$. The unperturbed Hamiltonian $H_0$
contains the total Hamiltonian given in Eq.~\eqref{eq:H_general}
except for the intersite hopping $H_{\text{hop}}^{i \neq j}$.  In the
limit $U_0 \gg t_{i\alpha,j\beta}$, we obtain, as it should be, the same
results with both methods.  In the region of physically relevant model
parameters $U_0$ and $J_{\text{avg}}$, there are nevertheless
higher-order corrections to the second order perturbation theory
results.

We now use spin projectors to obtain the sought effective spin
Hamiltonian.

\subsubsection*{Spin projectors}
Using the Abrikosov pseudo-fermion representation for  spin 1/2 operators, 
\begin{align*}
c_{i \uparrow}^\dagger c_{i \downarrow}=S_i^+, \qquad
c_{i \downarrow}^\dagger c_{i \uparrow}=S_i^-, \qquad \\
c_{i \uparrow}^\dagger c_{i \uparrow}=\frac{1}{2}+S_i^z, \qquad
c_{i \downarrow}^\dagger c_{i \downarrow}=\frac{1}{2}-S_i^z,
\end{align*}
and the fact that an operator in second quantization is expressed as
\begin{align*}
\hat{A}=\sum_{\mu \nu \mu^\prime \nu^\prime} \langle \mu \nu \vert \hat{A} 
\vert \mu^\prime \nu^\prime \rangle c_{1 \mu}^\dagger c_{2 \nu}^\dagger c_{2 \nu^\prime} c_{1 \mu^\prime},
\end{align*}
we can translate an effective spin-1/2 Hamiltonian written 
 in second quantization to a spin Hamiltonian.
 For example, the operator 
\begin{align*}
\vert \uparrow_{i,z^2} \uparrow_{j,z^2} \rangle \langle \uparrow_{i,z^2} \uparrow_{j,z^2} \vert 
&= c_{i,z^2 \uparrow}^\dagger c_{j,z^2 \uparrow}^\dagger c_{j,z^2 \uparrow} c_{i,z^2 \uparrow}
\end{align*}
leads to a term that couples the $z$ components of the spin
\begin{align*}
c_{i,z^2 \uparrow}^\dagger c_{j,z^2 \uparrow}^\dagger c_{j,z^2 \uparrow} c_{i,z^2 \uparrow}
= (\tfrac{1}{2}+S_{i}^z)(\tfrac{1}{2}+S_{j}^z).
\end{align*}
Having introduced spin-$1/2$ operators, we can recast the relevant
terms in the electronic Hamiltonian in the form of $3\times 3$
matrices that describe (anisotropic) interactions between the
components of the spins $1/2$ at sites $i$ and $j$.
The spin Hamiltonian then reads
\begin{align}
H_{\text{spin}} = \vec{S}_i^T \Gamma_{ij} \vec{S}_j
\label{Hspin}
\end{align}
where the bilinear spin-spin interaction matrix $\Gamma_{ij}$ has
components (see also Ref.~\onlinecite{Winter2016})
\begin{align}
\Gamma_{ij} = \begin{pmatrix}
\mathcal{J}_{ij}+\mathcal{K}_{ij}^{xx} & \mathcal{D}_{ij}^z + \mathcal{K}_{ij}^{xy} & -\mathcal{D}_{ij}^y + \mathcal{K}_{ij}^{xz} \\
-\mathcal{D}_{ij}^z +\mathcal{K}_{ij}^{xy} & \mathcal{J}_{ij}+\mathcal{K}_{ij}^{yy} & \mathcal{D}_{ij}^x + \mathcal{K}_{ij}^{yz} \\
\mathcal{D}_{ij}^y + \mathcal{K}_{ij}^{xz} & -\mathcal{D}_{ij}^x + \mathcal{K}_{ij}^{yz} & \mathcal{J}_{ij}-\mathcal{K}_{ij}^{xx}-\mathcal{K}_{ij}^{yy}
\end{pmatrix}.\label{gamma}
\end{align}
The matrix consists of the Heisenberg exchange $\mathcal{J}_{ij}$, the
Dzyaloshinskii-Moriya vector $\vec{\mathcal{D}}_{ij}$ and the
traceless symmetric tensor $\hat{\mathcal{K}}_{ij}$; the spin
Hamiltonian consequently has the form of Eq.~\eqref{eq:HSpin}. The
symmetric tensor $\hat{\mathcal{K}}_{ij}$ is chosen to be traceless to
ensure that the definition of the Heisenberg exchange
$\mathcal{J}_{ij}$ is not modified by considering additional
non-rotationally invariant terms.

We can now determine the values for the coupling parameters in
Eq.~(\ref{gamma}) from the previously derived {\it ab initio} hopping
parameters and $\lambda$. The details of the crystal structure which
influences the form of the exchange parameters, like the symmetry and
the orbital hierarchy are implicitly encoded in these {\it ab initio}
parameters.

\subsection{$\vec{\mathcal{D}}_{ij}$ and $\hat{\mathcal{K}}_{ij}$ in pyrochlore systems} \label{sec:Pyrochlore_symmetry}

Since $H_{\text{spin}}$ in Eq.~\eqref{Hspin} ought to be invariant
under the symmetry operations of the crystal, we analyze now the
symmetries of the pyrochlore lattice.  It is known from the Moriya
rules \cite{Moriya1960} that the direction of the
Dzyaloshinskii-Moriya vector $\vec{\mathcal{D}}_{ij}$ depends on
mirror planes and rotation axes in the system considered. In fact,
these symmetries also determine the number of independent parameters
in the symmetric tensor $\hat{\mathcal{K}}_{ij}$.

\begin{figure}
\includegraphics[width=0.48\textwidth]{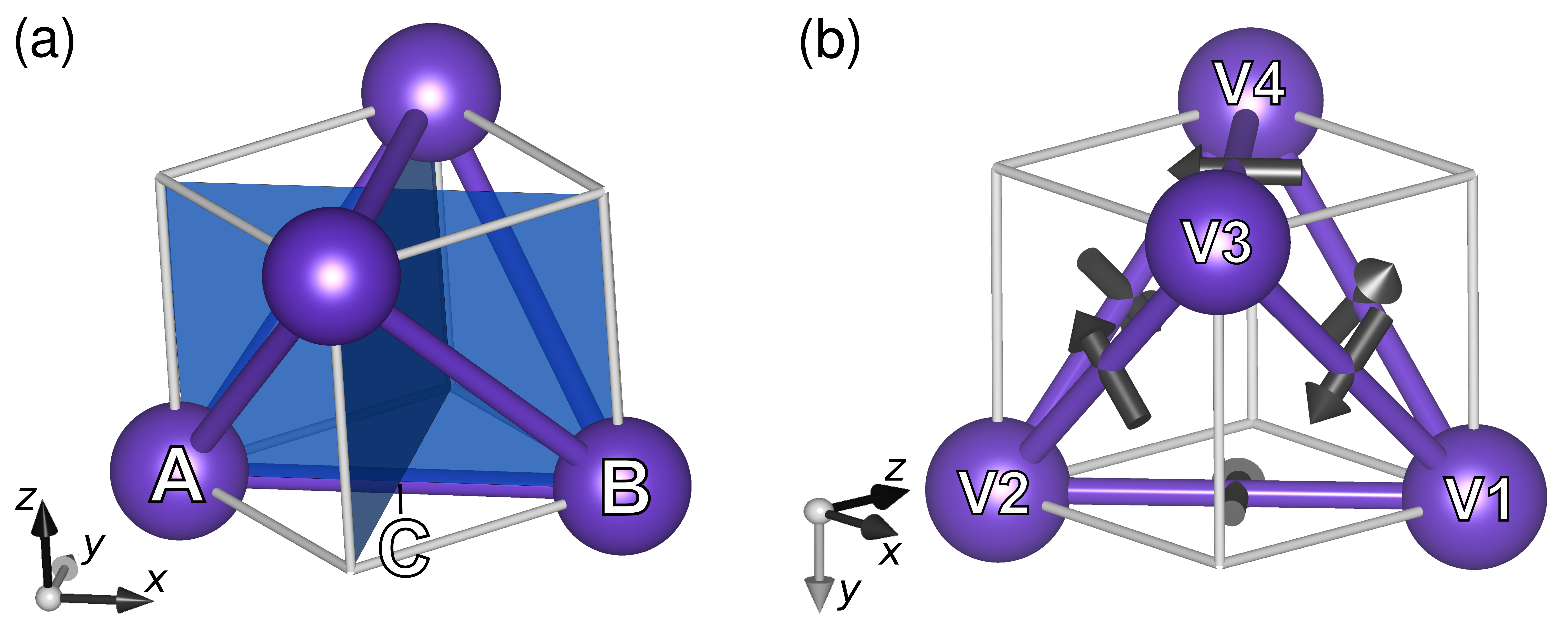}
\caption{(a) Vanadium tetrahedron showing the two mutually
  perpendicular mirror planes relevant to the A-B bond.  One plane
  includes the bond A-B, while the other bisects this bond at C.  The
  presence of such mirror planes constrains the form of the local DM
  vectors and symmetric tensors. (b) Direction of the
  Dzyaloshinskii-Moriya vectors in a pyrochlore system in the global
  coordinate system considered in the calculations.  }
\label{fig:Mirror_planes+DM}
\end{figure}

For simplicity, let us assume that one bond between sites A and B is
in the direction of the global $x$ axis, as shown in
Fig.~\ref{fig:Mirror_planes+DM}~(a). In the pyrochlore lattice, there
are two mirror planes which are important for the determination of the
symmetry properties of the exchange parameters.

One mirror plane is perpendicular to A-B and passes through C. Since
spin is a pseudovector, it transforms under this symmetry operation as
\begin{align*}
\begin{pmatrix}
S^x_A \\ S^y_A \\ S^z_A
\end{pmatrix} \rightarrow
\begin{pmatrix}
S^x_B \\ -S^y_B \\ -S^z_B
\end{pmatrix}, \qquad
\begin{pmatrix}
S^x_B \\ S^y_B \\ S^z_B
\end{pmatrix} \rightarrow
\begin{pmatrix}
S^x_A \\ -S^y_A \\ -S^z_A
\end{pmatrix},
\end{align*}
with the spin Hamiltonian having to be invariant under this symmetry
operation. Therefore, those terms in the Hamiltonian for which the
sign is changed under reflection have to vanish,
\begin{align*}
\mathcal{D}^x_{AB}=0, \qquad \mathcal{K}_{AB}^{xy}=0, \qquad \mathcal{K}_{AB}^{xz}=0.
\end{align*}
The second mirror plane includes A-B and lies in the $xz$ plane for
the chosen global coordinate system. With the symmetry operations
\begin{align*}
\begin{pmatrix}
S^x_A \\ S^y_A \\ S^z_A
\end{pmatrix} \rightarrow
\begin{pmatrix}
-S^x_B \\ S^y_B \\ -S^z_B
\end{pmatrix}, \qquad
\begin{pmatrix}
S^x_B \\ S^y_B \\ S^z_B
\end{pmatrix} \rightarrow
\begin{pmatrix}
-S^x_A \\ S^y_A \\ -S^z_A
\end{pmatrix}
\end{align*}
this leads to the restrictions
\begin{align*}
\mathcal{D}^x_{AB}=0, \qquad \mathcal{D}^z_{AB}=0, \qquad \mathcal{K}_{AB}^{xy}=0, \qquad \mathcal{K}_{AB}^{yz}=0.
\end{align*}
In conclusion, the direction of the Dzyaloshinskii-Moriya vector is,
except for its sign, fully determined by symmetry considerations. Its
only non-vanishing contribution is in the $y$ direction, perpendicular
to the bond under consideration and within the face of the cube
enclosing the tetrahedron. The symmetric tensor is diagonal for this
choice of coordinate system. We thus have
\begin{align*}
\vec{\mathcal{D}}_{AB}=\begin{pmatrix}
0 \\ \mathcal{D}^y_{AB} \\ 0
\end{pmatrix}, \,
\hat{\mathcal{K}}_{AB}=\begin{pmatrix}
\mathcal{K}^{xx}_{AB} & 0 & 0 \\
0 & \mathcal{K}^{yy}_{AB} & 0 \\
0 & 0 & -\mathcal{K}^{xx}_{AB}-\mathcal{K}^{yy}_{AB}
\end{pmatrix}.
\end{align*}

Hence, for the pyrochlore system with only nearest-neighbor
interactions, there is only one independent exchange parameter for the
Dzyaloshinskii-Moriya vector and there are two independent parameters
that characterize the symmetric tensor $\hat{ \mathcal{K}}_{ij}$.

In our calculations, we worked in a global coordinate system aligned
along the cubes edges.  We show in Fig.~\ref{fig:Mirror_planes+DM}~(b)
all the DM vectors for one tetrahedron, with their explicit form given
in Appendix~\ref{sec:DM_vectors} in this global coordinate system.  To
find the correct contributions to both the DM vector and the symmetric
tensor $\hat{\mathcal{K}}$ within this description, one has to rotate
the coordinate system used above.

As an example, we give the result for the bond 1-2, as defined in
Fig. \ref{fig:Mirror_planes+DM}~(b), which is obtained by a rotation
of $\pi$/4 about the $z$ axis and a rotation of $\pi$/2 about the $x$
axis
\begin{align*}
\vec{\mathcal{D}}_{12}=\begin{pmatrix}
\mathcal{D}_{12}^x \\ 0 \\ \mathcal{D}_{12}^x
\end{pmatrix}, \;
\hat{\mathcal{K}}_{12}=\begin{pmatrix}
\mathcal{K}^{xx}_{12} & 0 & \mathcal{K}^{xz}_{12} \\
0 & -2 \mathcal{K}^{xx}_{12} & 0 \\
\mathcal{K}^{xz}_{12} & 0 & \mathcal{K}^{xx}_{12}
\end{pmatrix},
\end{align*}
with $\mathcal{K}_{12}^{xx}=\frac{1}{2}(\mathcal{K}_{AB}^{xx}+\mathcal{K}_{AB}^{yy})$, 
$\mathcal{K}_{12}^{yy}=\frac{1}{2}(\mathcal{K}_{AB}^{xx}-\mathcal{K}_{AB}^{yy})$ and $\mathcal{D}_{12}^x = \frac{\mathcal{D}_{AB}^y}{\sqrt{2}}$.

As previously noted for odd electron ions in pyrochlore
systems\cite{Ross2011}, we therefore have, together with the isotropic
Heisenberg exchange $\mathcal{J}_{12}$, four independent bilinear
spin-spin couplings, in principle~\cite{Rau_BYZO}.  We discuss in
Appendix~\ref{CHOICE} various representations of equivalent spin
Hamiltonians using different spin quantization frames.

\begin{figure}
\includegraphics[width=0.43\textwidth]{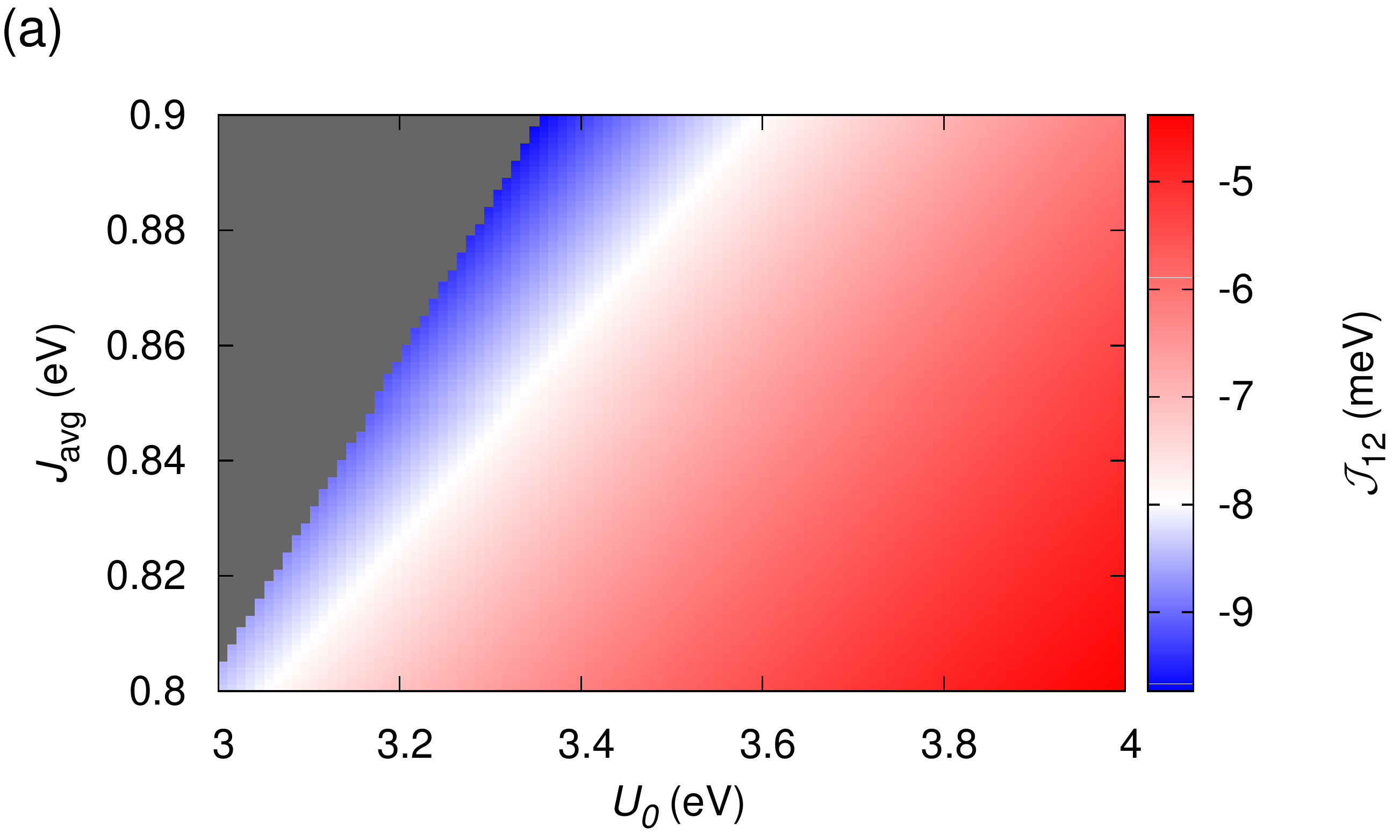}
\includegraphics[width=0.43\textwidth]{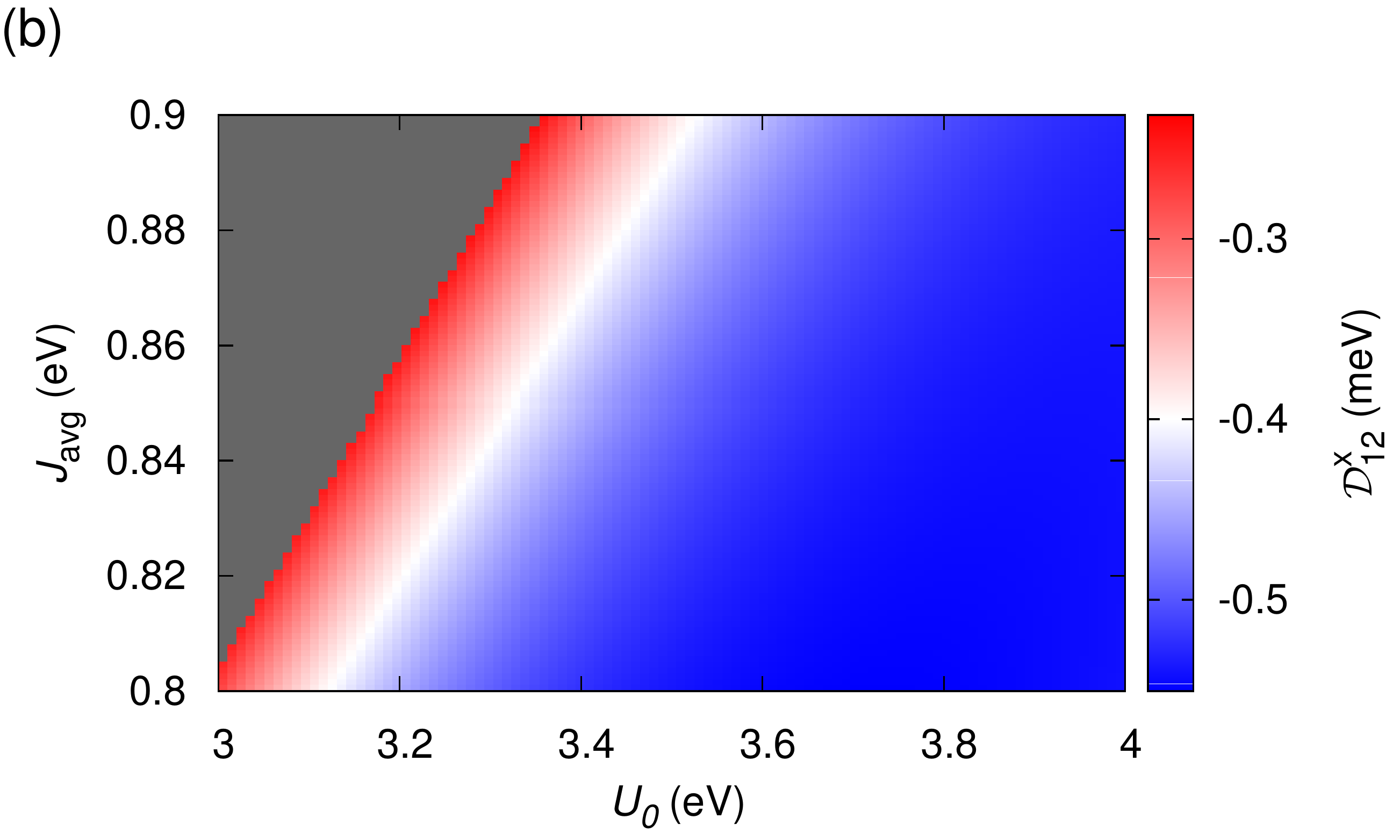}
\includegraphics[width=0.43\textwidth]{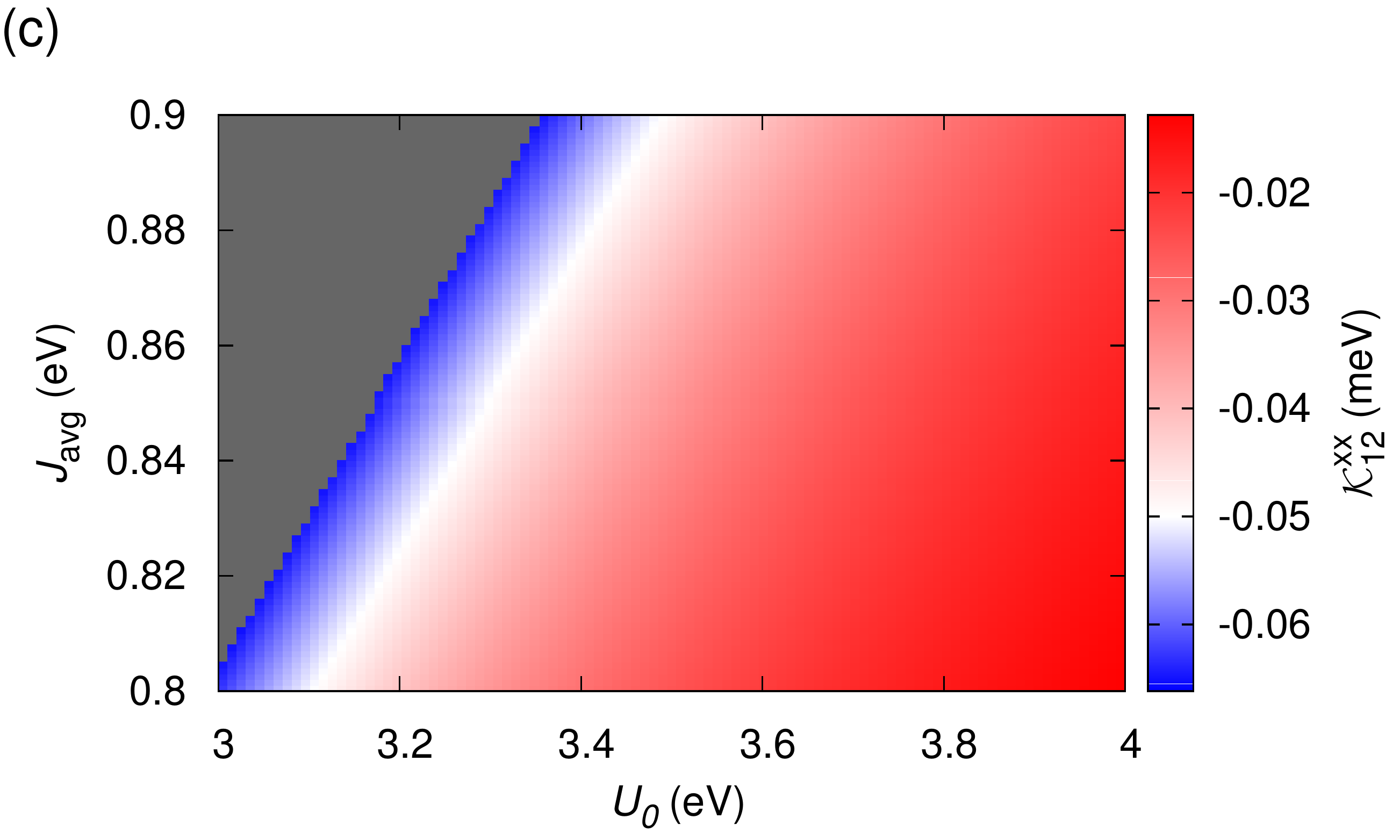}
\includegraphics[width=0.43\textwidth]{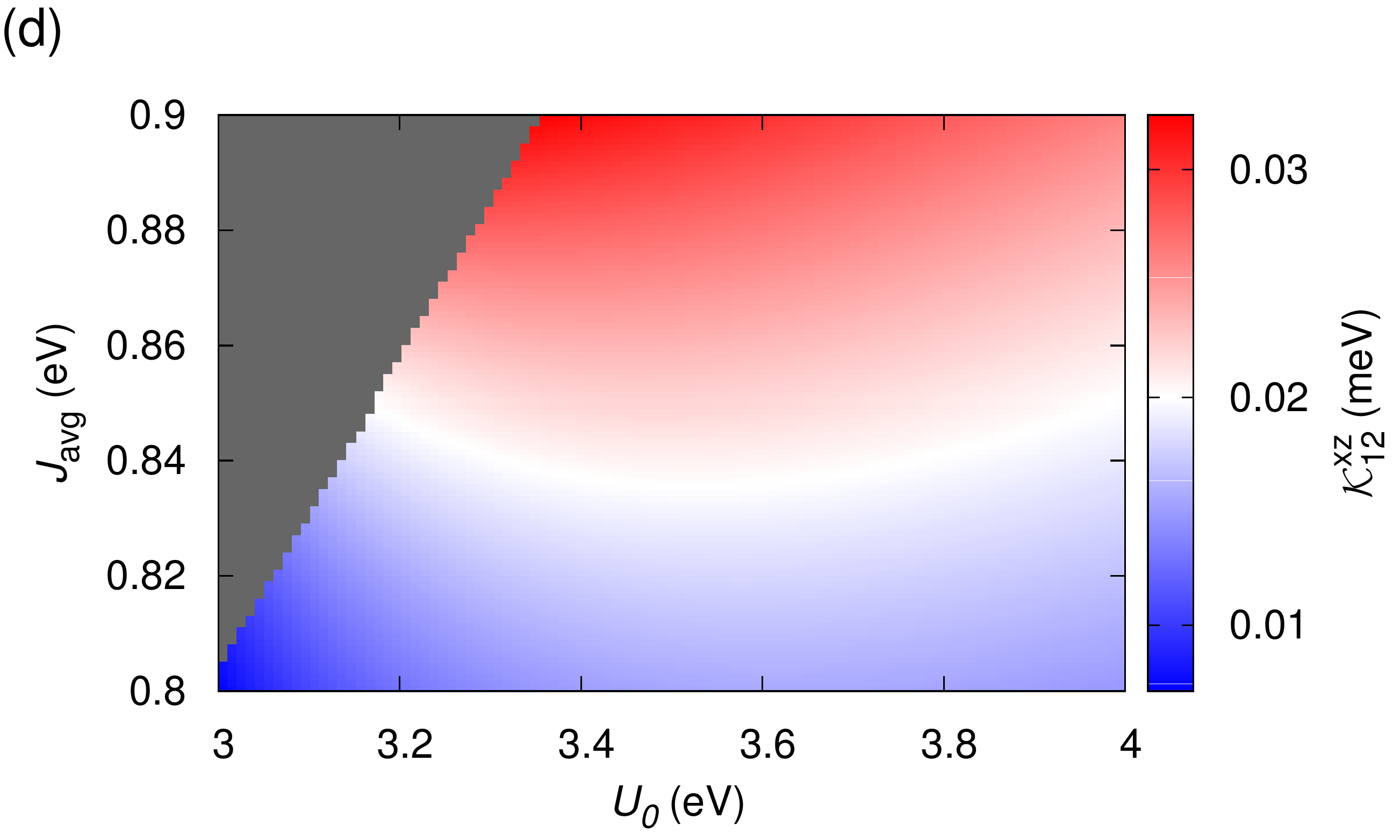}
\caption{Dependence of the calculated exchange parameters on the
  interaction parameters $U_0$ and $J_{\text{avg}}$. The results for a
  certain choice of interaction parameters are given in
  Eq.~\eqref{eq:Parameters12}. The gray area far left is the forbidden region
  where, for $U_0-3J_{\alpha \beta}<0$, the projection on singly
  occupied states is no longer justified and the white lines
  correspond to the theoretical results given in
  Eq.~\eqref{eq:Parameters12}. (a) Heisenberg exchange
  $\mathcal{J}_{12}$. For comparison, experiment~\cite{Mena2014}
  estimates $\vert \mathcal{J} \vert = 8.22$~meV. (b)
  $\mathcal{D}^x_{12}$, given in the global coordinate system as in
  Fig.~\ref{fig:Mirror_planes+DM}~(b). The results are below the
  experimental estimates which correspond to
  $\mathcal{D}^x_{12}=0.8$~meV~\cite{Onose2010} and
  $\mathcal{D}^x_{12}=1.05$~meV~\cite{Mena2014}. Note, that we are reporting 
  $\mathcal{D}^x_{12}$ and not $\vert \vec{ \mathcal{D}}_{12} \vert= \sqrt{2} \, \mathcal{D}^x_{12}$. (c), (d)
  Independent contributions to the symmetric anisotropic exchange
  $\hat{\mathcal{K}}$. The calculated values are for none of the interaction parameters $U_0$, $J_{\text{avg}}$ 
  small enough to neglect them.  }
\label{fig:Energy_parameters}
\end{figure}

The dependence of the energy parameters in the spin Hamiltonian on the
model parameters $U_0$ and $J_{\text{avg}}$ in Lu$_2$V$_2$O$_7$ is
shown in Fig.~\ref{fig:Energy_parameters}.  The Hund's coupling within
$3d$ orbitals~\cite{Khomskii2014} is estimated to be $0.8-0.9\text{
  eV}$. For the Coulomb repulsion $U_0$ on the same orbital on
V$^{4+}$ ions, we considered values between 3 and $4\text{ eV}$.

With the parameter choice $U_0=3.3\text{ eV}$ and
$J_{\text{avg}}=0.845\text{ eV}$, we extract from the effective
Hamiltonian (coefficients given in Table~\ref{tab:Coeff_ortho}) the
following energy parameters (all in meV)
\begin{align} \label{eq:Parameters12}
\mathcal{J}_{12}&=-7.99, \quad
\vec{\mathcal{D}}_{12} = \begin{pmatrix}
-0.4 \\ 0 \\ 0.4
\end{pmatrix}, \nonumber \\
\hat{\mathcal{K}}_{12} &= \begin{pmatrix}
-0.05 & 0 & 0.02 \\
0 & 0.1 & 0 \\
0.02 & 0 & -0.05
\end{pmatrix}.
\end{align}
The isotropic Heisenberg exchange $\mathcal{J}_{12}$ is ferromagnetic
and in reasonably good agreement with the experimental value
$\mathcal{J} = -8.22$~meV in Ref.~\onlinecite{Mena2014}, and not too
far off from the results obtained from \textit{ab initio} calculations
with the ``energy mapping method'' of Ref.~\onlinecite{Xiang2011}
($\mathcal{J}=-7.09$~meV). We decided to focus on a choice of
interaction parameters, based on the agreement of the Heisenberg
exchange with the inelastic neutron scattering
experiment~\cite{Mena2014}, so that our calculated values for the
Dzyaloshinskii-Moriya interaction and the symmetric tensor can be more
easily compared with the experimental results.

As explained above and as illustrated in
Fig.~\ref{fig:Mirror_planes+DM}~(b), the directions of the
$\vec{\mathcal{D}}_{ij}$ vectors are fully
determined\cite{Moriya1960,Elhajal2005}, with only its sign being
free.
The numerical results completely agree with the prediction of Moriya's
rules, which is a confirmation that the choice of local coordinate
systems and the rotations performed throughout our calculations implement the
crystal symmetry correctly.  The orientation of the DM vectors that we
find here corresponds to the so-called ``indirect'' case of
Ref.~\onlinecite{Elhajal2005}.
The ratio
\begin{equation}
\frac{\vert \vec{\mathcal{D}} \vert}{\vert \mathcal{J}\vert} \approx 0.07,
\label{DoverJ}
\end{equation}
where $\vert \vec{\mathcal{D}} \vert$ is calculated as
$\sqrt{\mathcal{D}_x^2+\mathcal{D}_y^2+\mathcal{D}_z^2}$, is lower
than the two experimental results~\cite{Onose2010,Mena2014} for this
material. By fitting transport data, Ref.~\onlinecite{Onose2010}
determined $\vert \vec{\mathcal{D}}/\mathcal{J}\vert \simeq 0.32$,
while the ratio obtained from inelastic neutron scattering
\cite{Mena2014} is $\vert \vec{\mathcal{D}}/\mathcal{J}\vert \simeq
0.18$.

The symmetric tensor results in smaller corrections to the
nearest-neighbor Heisenberg spin Hamiltonian, though these are {\it
  not} entirely negligible when compared to ${\vert \vec{\mathcal{D}}
  \vert}/{\vert \mathcal{J}\vert}$ in Eq.~(\ref{DoverJ}), with
\begin{align*}
\frac{\vert \hat{\mathcal{K}} \vert}{\vert \mathcal{J}\vert} = 0.02,
\end{align*}
where $\vert \hat{\mathcal{K}} \vert$ is the Frobenius norm of the
symmetric tensor.

\section{Summary}\label{DISCUSSION}

In summary, we presented a method to determine the spin exchange
parameters of a spin $1/2$ system combining non-relativistic and
relativistic \textit{ab initio} density functional theory (DFT)
calculations with exact diagonalization of a generalized Hubbard model
on a finite cluster.  Projecting the Hamiltonian onto the low energy
subspace and using spin projectors, we transformed the effective
Hamiltonian into a spin Hamiltonian, considering all allowed isotropic
and anisotropic nearest-neighbor bilinear exchange parameters for a
spin $1/2$ system.

We determined for Lu$_2$V$_2$O$_7$ the four independent exchange
parameters for a certain choice of Hubbard repulsion $U_0$ and Hund's
coupling $J_{\text{avg}}$.

The isotropic (Heisenberg) exchange parameter $\mathcal{J}_{ij}$ that
we determined, $\mathcal{J}_{ij}\approx -8\,$meV, is close to the
experimental value ($\mathcal{J}_{ij} \approx -8.22\,$meV) extracted
from inelastic neutron scattering (INS) data~\cite{Mena2014}.  On the
other hand, the ratio of the Dzyaloshinskii-Moriya interaction to the
Heisenberg exchange in Eq.~(\ref{DoverJ}) is almost a factor 3 smaller
than the $\vert \vec{\mathcal{D}}_{ij} \vert / \vert \mathcal{J}_{ij}
\vert \simeq 0.18$ obtained from the same INS data. As such, the
discrepancy between the present calculations and the $\vert
\vec{\mathcal{D}}_{ij} \vert / \vert \mathcal{J}_{ij} \vert$ value
extracted from INS appears rather large. We comment further on that
below.  We note that the $\mathcal{J}_{ij} \approx -3.4\,$meV value
found by fitting the magnetic specific heat\cite{stiffness} in
Ref.~\onlinecite{Onose2010} is significantly different from
both the INS value and the present DFT result. This may suggest a
necessity to reinvestigate the low-temperature magnetic specific heat
data of this compound as well as reanalyzing it.  In the same vein,
the $\vert \vec{\mathcal{D}}_{ij} \vert / \vert \mathcal{J}_{ij} \vert
\simeq 1/3$ determined by fitting transport data~\cite{Onose2010} is
significantly larger than both the present DFT ratio
(Eq.~\eqref{DoverJ}) and the INS results~\cite{Mena2014}.  It is
unclear to what extent this surprisingly large ratio for a $3d$
transition metal ion (V$^{4+}$) results from the small
$\mathcal{J}_{ij} \approx -3.4\,$meV found from specific heat\cite{Onose2010, stiffness}.  As the
INS data directly determines the spin-stiffness, there appears to be no
simple way in which the magnetic specific heat value $\mathcal{J}_{ij}
\approx -3.4\,$meV could be reconciled with the $\mathcal{J}_{ij}
\approx -8.22\,$meV value that parameterizes the spin stiffness directly
probed by INS.

Returning to the aforementioned difference between the INS and DFT
$\vert \vec{\mathcal{D}}_{ij} \vert / \vert \mathcal{J}_{ij} \vert$
ratios, a few comments are in order. First of all, the fit to the INS
data considered only nearest-neighbor exchange and 
Dzyaloshinskii-Moriya interactions, ignoring symmetric anisotropic
exchange (the two components of $\hat{\mathcal{K}}_{ij}$) as well as
any interaction beyond nearest-neighbors.  Incorporating those in the
fitting could lead to a renormalization of the $\vert
\vec{\mathcal{D}}_{ij} \vert / \vert \mathcal{J}_{ij} \vert$ ratio.
The reason being that the spin-stiffness, determined through the
quadratic momentum dependence of the magnon dispersion near the zone
center~\cite{Mena2014}, would no longer solely, and uniquely, fix
$\mathcal{J}_{ij}$.  It may also be worthwhile to explore the effect
of the subleading anisotropic components ($\hat{\mathcal{K}}_{ij}$) on
the dispersion.  Similarly, we only considered (two-site)
nearest-neighbor interactions in our calculations and, as such, the
accuracy of our exchange parameters are also hampered by the same
distance truncation of the spin Hamiltonian used in the INS data
analysis.

At first sight, it would naively appear that the insulating
Lu$_2$V$_2$O$_7$ pyrochlore, with its ferromagnetic ground state and a
transition temperature of $T_c\approx 70$ K, should be a textbook
example of spin-$1/2$ ferromagnetism on a non-Bravais lattice well
described by isotropic Heisenberg exchange with leading anisotropic
exchange perturbations in the form of Dzyaloshinskii-Moriya (DM)
interactions.  However, at the present time, there appears to be some
significant discrepancy between the scale of the DM interaction
determined from transport measurements~\cite{Onose2010}, inelastic
neutron scattering data~\cite{Mena2014} and, from our density
functional theory calculations.  It would certainly be comforting, in
terms of one's understanding of what would appear as ``simple''
ferromagnetism of localized moments, to resolve this disagreement.
Furthermore, in view of the interests in topological aspects of magnon
excitations induced by antisymmetric spin-spin
interactions~\cite{Onose2010,Hirschberger,Chisnell}, a definite
progress in obtaining a quantitative global understanding of the
magnetic properties of Lu$_2$V$_2$O$_7$ would be useful.

\begin{acknowledgments}
  The authors acknowledge fruitful discussions with 
Steffen Backes,  
Vladislav Borisov,
Zhihao Hao, 
Claudine Lacroix,
Paul McClarty, 
Jeff Rau,
Igor Solovyev, 
Julian Stobbe,
Igor Mazin, 
Stephen Winter
and 
Alexander Yaresko.
The work at Frankfurt U. was supported by the Deutsche
Forschungsgemeinschaft (DFG) through project SFB/TRR49. The work at
the U. of Waterloo was supported by the NSERC of Canada, the Canada
Research Chair program (M.J.P.G., Tier 1), the Canadian Foundation for
Advanced Research and the Perimeter Institute (PI) for Theoretical
Physics.  Research at PI is supported by the Government of Canada
through Industry Canada and by the Province of Ontario through the
Ministry of Economic Development \& Innovation.

\end{acknowledgments}

\appendix
\section{Choice of spin Hamiltonian representation}\label{CHOICE}

We discuss here the various representations of the generalized
bilinear anisotropic Hamiltonian for spins $1/2$ on the pyrochlore
lattice.  At least three different ways to parameterize the
nearest-neighbor spin Hamiltonian have been employed.

In our study, we have used Eq.~\eqref{eq:HSpin}, which, due to its
general form, is not limited to the description of pyrochlores, but
applicable to any other crystal symmetry. However, as it does not
explicitly expose the relevant symmetries of the specific system
considered, it has the disadvantage that it seems to have more free
parameters than there actually are. Consequently, one has to introduce
additional symmetry considerations like those presented in
Section~\ref{sec:Pyrochlore_symmetry}.

A more specific choice of parameterization for the case of pyrochlores
was used by Thompson {\it et al.}~\cite{Thompson2011} with
four different nearest-neighbor bilinear exchange interactions
\begin{align}
H_{\text{ex}} = H_{\text{Ising}} + H_{\text{iso}} + H_{\text{pd}} + H_{\text{DM}}.
\end{align}
It contains an Ising like term with the spin projection on the local
$z$ axes on the site respectively
\begin{align}
H_{\text{Ising}}=-\mathcal{J}_{\text{Ising}} \sum_{\langle ij \rangle}  (\vec{S}_i \cdot \hat{z}_i ) (\vec{S}_j \cdot \hat{z}_j ),
\end{align}
an isotropic term which has Heisenberg character,
\begin{align}
H_{\text{iso}}=-\mathcal{J}_{\text{iso}} \sum_{\langle ij \rangle} \vec{S}_i \cdot \vec{S}_j,
\end{align}
a pseudo-dipolar term with projection on the bond $\hat{r}_{ij}$
connecting site $i$ and $j$,
\begin{align}
H_{\text{pd}}&=-\mathcal{J}_{\text{pd}}  \sum_{\langle ij \rangle}  (\vec{S}_i \cdot \vec{S}_j - 3 (\vec{S}_i \cdot \hat{r}_{ij} ) ( \vec{S}_j \cdot \hat{r}_{ij}) ),
\end{align}
and a term which was  labeled as the Dzyaloshinskii-Moriya term
\begin{align}
H_{\text{DM}}&=-\mathcal{J}_{\text{DM}} \, \vec{\Omega}_{\text{DM}}^{ij} \cdot ( \vec{S}_i \times \vec{S}_j ).
\end{align}
This way of parameterization includes only four different exchange
parameters and is therefore convenient to describe a pyrochlore
system, as shown in the previous section. By introducing the
Ising-like term, there are additional contributions of the
Dzyaloshinskii-Moriya term, $\vert \vec{\mathcal{D}}_{ij} \vert=
-\mathcal{J}_{\text{DM}}^{ij}-\frac{\sqrt{2}}{3}\mathcal{J}_{\text{Ising}}^{ij}$,
so that $\mathcal{J}_{\text{DM}}^{ij}$ and $\vert
\vec{\mathcal{D}}_{ij} \vert$ describe the strength of different
exchange processes. For the bond 1-2, the relation between
parameterizations is
\begin{align}
\mathcal{J}_{\text{Ising}}^{12} &= 9 \mathcal{K}_{12}^{xx} - 3 \mathcal{K}_{12}^{xz}, \\
\mathcal{J}_{\text{iso}}^{12} &= -\mathcal{J}_{12} + \mathcal{K}_{12}^{xx} - \frac{1}{3} \mathcal{K}_{12}^{xz}, \\
\mathcal{J}_{\text{pd}}^{12} &= -2 \mathcal{K}_{12}^{xx} + \frac{4}{3} \mathcal{K}_{12}^{xz}, \\
\mathcal{J}_{\text{DM}}^{12} &= -\vert \vec{\mathcal{D}}_{12} \vert - 3 \sqrt{2} \mathcal{K}_{12}^{xx}+\sqrt{2} \mathcal{K}_{12}^{xz}.
\end{align}
The relation for the other bonds can be obtained by considering the
pyrochlore symmetry.

Moreover, there is a third popular way of parameterization, introduced
in Ref.~\onlinecite{Ross2011} which, in an appendix, already pointed
out the difference with the formalism used in
Ref.~\onlinecite{Thompson2011}.  In Eq.~(2) of
Ref.~\onlinecite{Ross2011}, the parameter matrix is explicitly given
for a bond, which corresponds to bond 1-3 in
Fig.~\ref{fig:Mirror_planes+DM}~(b),
\begin{align}
\mathcal{J}_{\text{par}} = \begin{pmatrix}
\mathcal{J}_2 & \mathcal{J}_4 & \mathcal{J}_4 \\
-\mathcal{J}_4 & \mathcal{J}_1 & \mathcal{J}_3 \\
-\mathcal{J}_4 & \mathcal{J}_3 & \mathcal{J}_1
\end{pmatrix}.
\end{align}
This leads to a modification of the Heisenberg exchange used in our
notation and a renormalization of the Dzyaloshinskii-Moriya parameter,
\begin{align}
\mathcal{J}_1 &= \mathcal{J}_{13} - \frac{1}{2} \mathcal{K}_{13}^{xx}, \\
\mathcal{J}_2 &= \mathcal{J}_{13} + \mathcal{K}_{13}^{xx}, \\
\mathcal{J}_3 &= \mathcal{K}_{13}^{yz}, \\
\mathcal{J}_4 &= - \frac{1}{\sqrt{2}} \vert \vec{\mathcal{D}}_{13} \vert.
\end{align}

\section{Interaction parameters in H$_{\text{int}}$} \label{sec:H_int_parameters}
We use the definition for the orbital dependent Coulomb repulsion as used in Ref.~\onlinecite{Pavarini2011}.

\medskip
\begin{tabular}{lccccc}
$J_{\alpha \beta}$	& $\vert d_{x^2\text{-}y^2} \rangle$ & $\vert d_{z^2} \rangle$ & $\vert d_{xy} \rangle$ & $\vert d_{yz} \rangle$ & $\vert d_{xz} \rangle$ \\
$\vert d_{x^2\text{-}y^2} \rangle$	& $U_0$ 	& $j_2$ 	& $j_3$ 	& $j_1$ 	& $j_1$ \\
$\vert d_{z^2} \rangle$				& $j_2$ 	& $U_0$		& $j_2$ 	& $j_4$ 	& $j_4$ \\
$\vert d_{xy} \rangle$				& $j_3$ 	& $j_2$ 	& $U_0$ 	& $j_1$ 	& $j_1$ \\
$\vert d_{yz} \rangle$				& $j_1$ 	& $j_4$ 	& $j_1$ 	& $U_0$ 	& $j_1$ \\
$\vert d_{xz} \rangle$				& $j_1$ 	& $j_4$ 	& $j_1$ 	& $j_1$ 	& $U_0$
\end{tabular}
\medskip

The interaction parameters $j_n$ can be expressed in terms of the
Slater integrals~\cite{Liechtenstein95} $F_k$ as
follows~\cite{Pavarini2011}:
\begin{align}
j_1 &= \frac{3}{49} F_2 + \frac{20}{9} \frac{1}{49} F_4 \\
j_2 &= -2 J_{\text{avg}} + 3 j_1 \\
j_3 &= 6 J_{\text{avg}} - 5 j_1 \\
j_4 &= 4 J_{\text{avg}} - 3 j_1,
\end{align}
and where
\begin{align}
U_0 = F_0 + \frac{8}{5} J_{\text{avg}},  \quad \text{and} \quad
J_{\text{avg}} = \frac{5}{7}\frac{( F_2 + F_4 )}{14},
\end{align}
with $F_4=\frac{5}{8}F_2$. In this work, we choose as free independent
parameters $U_0$ and $J_{\text{avg}}$.  The Coulomb repulsion matrix
used in the interaction Hamiltonian~\eqref{H_int} can be easily
constructed as $U_{\alpha \beta}=3 \, U_0 \mathbbm{1} - 2 J_{\alpha
  \beta}$.

\section{Local coordinate system} \label{sec:local_coordinates} For a
unit cell with the vanadium positions as follows (used within our
calculations with full-potential local orbital (FPLO) basis),
\begin{align} \label{eq:V_positions}
\vec{v}_1&=\begin{pmatrix} 1/2 \\ 1/2 \\ 1/2 \end{pmatrix}, \;
\vec{v}_2=\begin{pmatrix} 1/4 \\ 1/2 \\ 1/4 \end{pmatrix}, \nonumber \\
\vec{v}_3&=\begin{pmatrix} 1/2 \\ 1/4 \\ 1/4 \end{pmatrix}, \;
\vec{v}_4=\begin{pmatrix} 1/4 \\ 1/4 \\ 1/2 \end{pmatrix}.
\end{align}
we use the local coordinate systems
\begin{alignat*}{8}
\vec{x}_1&=\frac{1}{\sqrt{2}}\begin{pmatrix} 0 \\ 1 \\ -1 \end{pmatrix}, \;
&&\vec{y}_1&&=\frac{1}{\sqrt{6}}\begin{pmatrix} -2 \\ 1 \\ 1 \end{pmatrix}, \;
&&\vec{z}_1&&=\frac{1}{\sqrt{3}}\begin{pmatrix} 1 \\ 1 \\ 1 \end{pmatrix}, \\
\vec{x}_2&=\frac{1}{\sqrt{2}}\begin{pmatrix} 0 \\ 1 \\ 1 \end{pmatrix}, \;
&&\vec{y}_2&&=\frac{1}{\sqrt{6}}\begin{pmatrix} 2 \\ 1 \\ -1 \end{pmatrix}, \;
&&\vec{z}_2&&=\frac{1}{\sqrt{3}}\begin{pmatrix} -1 \\ 1 \\ -1 \end{pmatrix}, \\
\vec{x}_3&=\frac{1}{\sqrt{2}}\begin{pmatrix} 0 \\ -1 \\ 1 \end{pmatrix}, \;
&&\vec{y}_3&&=\frac{1}{\sqrt{6}}\begin{pmatrix} -2 \\ -1 \\ -1 \end{pmatrix}, \;
&&\vec{z}_3&&=\frac{1}{\sqrt{3}}\begin{pmatrix} 1 \\ -1 \\ -1 \end{pmatrix}, \\
\vec{x}_4&=\frac{1}{\sqrt{2}}\begin{pmatrix} 0 \\ -1 \\ -1 \end{pmatrix}, \; 
&&\vec{y}_4&&=\frac{1}{\sqrt{6}}\begin{pmatrix} 2 \\ -1 \\ 1 \end{pmatrix}, \; 
&&\vec{z}_4&&=\frac{1}{\sqrt{3}}\begin{pmatrix} -1 \\ -1 \\ 1 \end{pmatrix}.
\end{alignat*}
Within these local coordinate systems and for the vanadium atoms at
the given positions, we obtain onsite energies as given in
Table~\ref{tab:Hop_onsite}, the most important hopping parameter
between nearest neighbors are given in Table~\ref{tab:Hop_NN}.
\begin{table} 
\begin{tabular}{l|rrrrr}
 & V1 $d_{x^2\text{-}y^2}$ & V1 $d_{z^2}$ & V1 $d_{xy}$ & V1 $d_{yz}$ & V1 $d_{xz}$ \\
\hline
V1 $d_{x^2\text{-}y^2}$ & 1.5815 	& 0			& 0		 	& -1.2612 		& 0 \\
V1 $d_{z^2}$ 			& 0			& 0.2351	& 0			& 0				& 0 \\
V1 $d_{xy}$ 			& 0			& 0			& 1.5815	& 0				& -1.2612 \\
V1 $d_{yz}$ 			& -1.2612	& 0			& 0			& 1.8316		& 0 \\
V1 $d_{xz}$ 			& 0			& 0			& -1.2612	& 0				& 1.8316
\end{tabular}
\caption{Onsite energies $t_{1\alpha,1\beta}$ (in eV) on vanadium site No.~1, the other three vanadium ions are symmetry equivalent.} \label{tab:Hop_onsite}
\end{table}
\begin{table} 
\begin{tabular}{l|rrrrr}
 & V2 $d_{x^2\text{-}y^2}$ & V2 $d_{z^2}$ & V2 $d_{xy}$ & V2 $d_{yz}$ & V2 $d_{xz}$ \\
\hline
V1 $d_{x^2\text{-}y^2}$ & -0.1384 	& 0.0710 	& -0.0916 	& -0.0088	& 0.1113 \\
V1 $d_{z^2}$ 			& 0.0710	& -0.0421	& 0.1229	& -0.0869	& -0.1506 \\
V1 $d_{xy}$ 			& -0.0916	& 0.1229	& -0.2441	& 0.1113	& 0.1198 \\
V1 $d_{yz}$ 			& -0.0088	& -0.0869	& 0.1113	& -0.0348	& 0.0383 \\
V1 $d_{xz}$ 			& 0.1113	& -0.1506	& 0.1198	& 0.0383	& 0.0093
\end{tabular}

\bigskip
\begin{tabular}{l|rrrrr}
 & V3 $d_{x^2\text{-}y^2}$ & V3 $d_{z^2}$ & V3 $d_{xy}$ & V3 $d_{yz}$ & V3 $d_{xz}$ \\
\hline
V1 $d_{x^2\text{-}y^2}$ & -0.2970 	& -0.1419 	& 0		 	& 0.1840 		& 0 \\
V1 $d_{z^2}$ 			& -0.1419	& -0.0421	& 0			& 0.1738		& 0 \\
V1 $d_{xy}$ 			& 0			& 0			& -0.0855	& 0				& -0.0731 \\
V1 $d_{yz}$ 			& 0.1840	& 0.1738	& 0			& 0.0314		& 0 \\
V1 $d_{xz}$ 			& 0			& 0			& -0.0731	& 0				& -0.0569
\end{tabular}

\bigskip
\begin{tabular}{l|rrrrr}
 & V4 $d_{x^2\text{-}y^2}$ & V4 $d_{z^2}$ & V4 $d_{xy}$ & V4 $d_{yz}$ & V4 $d_{xz}$ \\
\hline
V1 $d_{x^2\text{-}y^2}$ & -0.1384 	& 0.0710 	& 0.0916 	& -0.0088	& -0.1113 \\
V1 $d_{z^2}$ 			& 0.0710	& -0.0421	& -0.1229	& -0.0869	& 0.1506 \\
V1 $d_{xy}$ 			& 0.0916	& -0.1229	& -0.2441	& -0.1113	& 0.1198 \\
V1 $d_{yz}$ 			& -0.088	& -0.0869	& -0.1113	& -0.0348	& -0.0383 \\
V1 $d_{xz}$ 			& -0.1113	& 0.1506	& 0.1198	& -0.0383	& 0.0093
\end{tabular}
\caption{Dominant hopping parameters $t_{1\alpha,j\beta}$ (in eV) between nearest neighbors always with respect to vanadium site No.~1, 
the other hopping parameters result from symmetry operations.} \label{tab:Hop_NN}
\end{table}

\section{Direction of Dzyaloshinskii-Moriya vectors} \label{sec:DM_vectors}
For a primitive unit cell with the basis coordinates as in Eq. (\ref{eq:V_positions}), we have for the direction of the DM vectors
\begin{align}
\hat{d}_{12}&=\frac{1}{\sqrt{2}} \begin{pmatrix} -1 \\ 0 \\ 1 \end{pmatrix}, \;
\hat{d}_{13}=\frac{1}{\sqrt{2}} \begin{pmatrix} 0 \\ 1 \\ -1 \end{pmatrix}, \nonumber \\
\hat{d}_{14}&=\frac{1}{\sqrt{2}} \begin{pmatrix} 1 \\ -1 \\ 0 \end{pmatrix}, \;
\hat{d}_{23}=\frac{1}{\sqrt{2}} \begin{pmatrix} -1 \\ -1 \\ 0 \end{pmatrix}, \nonumber \\
\hat{d}_{24}&=\frac{1}{\sqrt{2}} \begin{pmatrix} 0 \\ 1 \\ 1 \end{pmatrix}, \;
\hat{d}_{34}=\frac{1}{\sqrt{2}} \begin{pmatrix} -1 \\ 0 \\ -1 \end{pmatrix}.
\end{align}
This corresponds to the ``indirect'' case discussed in Ref.~\onlinecite{Elhajal2005}.

\bibliographystyle{apsrev4-1}

\begin{thebibliography}{99}
\bibitem{ring_exchange} We are implicitly considering here the lowest-order nontrivial hopping processes and not
discussing high-order ones giving rise, for example,  to so-called ring (or cyclic) exchange.
\bibitem{stiffness} Ref.~\protect\onlinecite{Onose2010} estimated a spin
  stiffness $D_s=-21\,\text{meV}\cdot\text{\AA}^2$.  The superexchange
  ${\mathcal{J}}=-3.4\,$meV can be obtained from the relationship
  ${\mathcal{J}}S=8D_s/a^2$ where $S=1/2$ is the spin value and
  $a=9.94\,${\AA} is the size of the conventional cubic cell of
  Lu$_2$V$_2$O$_7$.
\bibitem{Footnote:orbidx} Note that, although we chose for the sake of
  brevity, the labels $\alpha$, $\beta$ for the orbitals, it would be
  more correct to label them with an additional site index
  (e.g. $\alpha_i$, $\beta_i$) since the orbitals are described in
  real space and therefore depend, due to the different local
  coordinate systems, on the specific site under consideration.
\end{thebibliography}


\end{document}